\documentclass[a4paper,twocolumn,superscriptaddress,accepted=2023-02-13]{quantumarticle}
\pdfoutput=1
\usepackage[utf8]{inputenc}
\usepackage{amsfonts}
\usepackage{amsmath}
\usepackage{amsthm}
\usepackage{braket}
\usepackage{graphicx}
\usepackage{hyperref}
\usepackage[dvipsnames]{xcolor}
\usepackage{amssymb}
\usepackage{dsfont}
\usepackage{verbatim}
\usepackage{amsmath,amscd}
\usepackage{stmaryrd}
\usepackage{outlines}
\usepackage[numbers,sort&compress]{natbib}

\definecolor{dom}{RGB}{34,139,34}

\begin{document}

\title{A cellular automaton decoder for a noise-bias tailored color code}
\author{Jonathan F. San Miguel}
\affiliation{Department of Physics, Stanford University, Stanford, CA 94305}

\author{Dominic J. Williamson}
\affiliation{Department of Physics, Stanford University, Stanford, CA 94305}
\affiliation{Centre for Engineered Quantum Systems, School of Physics, University of Sydney, Sydney, NSW 2006, Australia}

\author{Benjamin J. Brown}
\affiliation{Centre for Engineered Quantum Systems, School of Physics, University of Sydney, Sydney, NSW 2006, Australia}

\begin{abstract}
    Self-correcting quantum memories demonstrate robust properties that can be exploited to improve active quantum error-correction protocols. Here we propose a cellular automaton decoder for a variation of the color code where the bases of the physical qubits are locally rotated, which we call the XYZ color code. The local transformation means our decoder demonstrates key properties of a two-dimensional fractal code if the noise acting on the system is infinitely biased towards dephasing, namely, no string-like logical operators. As such, in the high-bias limit, our local decoder reproduces the behavior of a partially self-correcting memory. At low error rates, our simulations show that the memory time diverges polynomially with system size without intervention from a global decoder, up to some critical system size that grows as the error rate is lowered. Furthermore, although we find that we cannot reproduce partially self-correcting behavior at finite bias, our numerics demonstrate improved memory times at realistic noise biases. Our results therefore motivate the design of tailored cellular automaton decoders that help to reduce the bandwidth demands of global decoding for realistic noise models.

\end{abstract}

\maketitle

\section{Introduction}

It remains a significant challenge to produce a large-scale quantum computer with noisy quantum systems. We therefore look for robust quantum error-correcting codes that can compensate for the limitations of quantum computing hardware as it is scaled to solve difficult problems reliably. Topological quantum error-correcting codes now make up the bedrock of modern designs to realize a scalable quantum computer. Notably, a significant majority of experimental efforts\cite{Linke17,Takita17,Andersen20,GoogleQuantumAI21,Egan21} are now working to produce a surface code\cite{Kitaev03, dennis2002, Raussendorf07, Fowler12}. The underlying physics of this code is a two-dimensional topological phase of matter that gives rise to anyonic quasiparticle excitations~\cite{Kitaev03}. Other phases, known as fracton topological phases~\cite{Chamon2005, haah2011, Vijay2016,Nandkishore2019,Pretko2020}, demonstrate exotic quasiparticle excitations with restricted dynamics that can be exploited to produce high-performance error-correcting codes~\cite{Brown2020, Nixon2021, Song2021}. 
In particular, so called type-II fracton topological phases~\cite{haah2011, Vijay2016} give rise to passive memories at finite temperature that demonstrate partial self correction~\cite{haah2011energy, haah2013, brown2016}.

Real physical qubits are likely to experience biases in their noise parameters. Certain qubits have even been designed to maintain their bias as they undergo unitary operations~\cite{mirrahimi2014dynamically,Puri2017, Guillaud2019, Puri2020, Grimm2020}. As such, considerable work has been invested in producing tailored quantum error-correcting codes together with specialized decoding algorithms that concentrate on correcting common types of error~\cite{aliferis_fault-tolerant_2008,aliferis_fault-tolerant_2009,Brooks13,Stephens13,Darmawan2017, Xu18, Tuckett2018, Darmawan18,Li2019,Nickerson19,Tuckett19,Tuckett2020, bonilla2021,guillaud2021error, darmawan2021, Higgott2021, srivastava2021,dua2022,higgott2022}. In the limit of very high bias, certain codes have been shown to reproduce the restricted dynamics of the quasiparticle excitations of fracton topological codes in lower-dimensional systems~\cite{Tuckett2020, bonilla2021, srivastava2021}. These dynamics can be understood in terms of the materialized symmetries~\cite{Kitaev03, Brown2020}, or more generally the system symmetries of a code together with its noise model~\cite{Tuckett2020}. Such symmetry restricted fractons have recently gained interest from a condensed matter perspective~\cite{Stephen2022} as they circumvent known restrictions on the existence of fractons in two dimensions~\cite{aasen2020}. 

Here, we investigate the dynamics of a variation of the color code, which we call the XYZ color code, that is modified to change the nature of its excitations. At infinite bias, this code reproduces the behavior of a type-II fracton code\cite{haah2013,PhysRevB.95.155133}, whose logical operators have a fractal-like support. In contrast, other examples such as the XZZX code~\cite{Wen03exact,bonilla2021} and the tailored surface code~\cite{Tuckett2018} give rise to lineons~\cite{bonilla2021} and type-I fractons~\cite{Tuckett2020, Nixon2021}, respectively, under an infinite bias noise model.

The generic features of passive self-correcting quantum memories can be exploited for active quantum error correction protocols where syndromes are measured and corrected manually. For instance, their corresponding codes can be readily decoded using local cellular automata\cite{bennett1985}. Such decoders\cite{dennis2002, Harrington04, Pastawski11, Hastings14, Herold2015, Breuckmann16, Herold17} are valuable when communication speed is limited. At a practical level, there is a limited bandwidth between quantum hardware that runs at milli-Kelvin temperatures~\cite{Reilly2019, Das2020, Delfosse2020}, and the classical control software running decoding algorithms, that operates outside of the dilution refrigerator.  More fundamentally, as we scale a quantum computer, the speed of light may limit its performance if the system depends on results from a global decoder that must first receive information from many non-local sites of a large system.

It is difficult to find practical quantum error-correcting codes that can be decoded locally. This is because known self-correcting memories are local only in four dimensions~\cite{dennis2002, Pastawski11, Hastings14, Breuckmann16}, and therefore require non-local interactions to be realized in three spatial dimensions. In contrast, work to produce cellular automaton decoders for two-dimensional codes under general noise models tend to compromise the threshold of the system~\cite{Harrington04, Herold2015, Breuckmann16, Herold17}. Moreover, known examples typically require some parameter, such as their speed or memory size, to diverge as the size of the quantum error-correcting code approaches infinity. Given these challenges it is interesting to design local decoders for low-dimensional codes under more constrained settings, such as highly biased noise models.

In this work we propose a cellular automaton decoder for the XYZ color code that is designed to mimic the behavior of partial self correction at infinite noise bias. To test the decoder, we conduct numerical simulations to measure how long the cellular automaton decoder can preserve the quantum memory without intervention from a global decoding algorithm. Like a partially self-correcting quantum memory, we find that the memory time diverges polynomially up to some fixed system size that depends on the rate of errors that the physical qubits experience. This means that if we aim to maintain a logical qubit for an arbitrarily long time, then the number of calls that need to be made to a global decoder decreases with system size, provided the physical error rate is sufficiently low. 

Furthermore, we also test our cellular automaton decoder at finite bias. Although we find that we cannot reproduce the standard signatures of partial self correction at any finite bias that is not diverging towards an infinitely biased noise model, we do find constant factor improvements in memory time. This amounts to approximately a factor of three increase in memory time for experimentally realistic biases, where the dephasing error rate is around 100-times greater than other noise processes. Our results therefore demonstrate that certain codes, together with local cellular automaton decoders, can be tailored to correct for biased noise to ease the demand placed on global decoding subroutines.

The remainder of the manuscript is structured as follows. In Sec.~\ref{sec:colorcode}, we introduce the XYZ color code, and describe an efficient maximum likelihood global decoder at infinite bias. Next, in Sec.~\ref{sec:cadecoder}, we describe the cellular automaton decoder, and derive its mapping to the Newman-Moore Model, a classical spin Hamiltonian with fracton-like excitations. Sec.~\ref{sec:infiniteresults} contains our numerical results at infinite bias. In Sec.~\ref{sec:finitebias}, we show results at finite bias, as well as describing the global renormalization-group decoder\cite{haah2013}, that we use as a benchmark. Finally, in Sec.~\ref{sec:conclusion}, we conclude and discuss future work. In Appendix \ref{app:optimalsize}, we show that there exists an infinite family of codes at appropriate system sizes with no degeneracy of logical operators at infinite bias, and in Appendix \ref{app:proofthreshold}, we prove a lemma that implies that the maximum likelihood decoder has a 50\% threshold at these system sizes.

\section{The XYZ Color Code}
\label{sec:colorcode}

\begin{figure}
    \centering
    \includegraphics[width=\columnwidth]{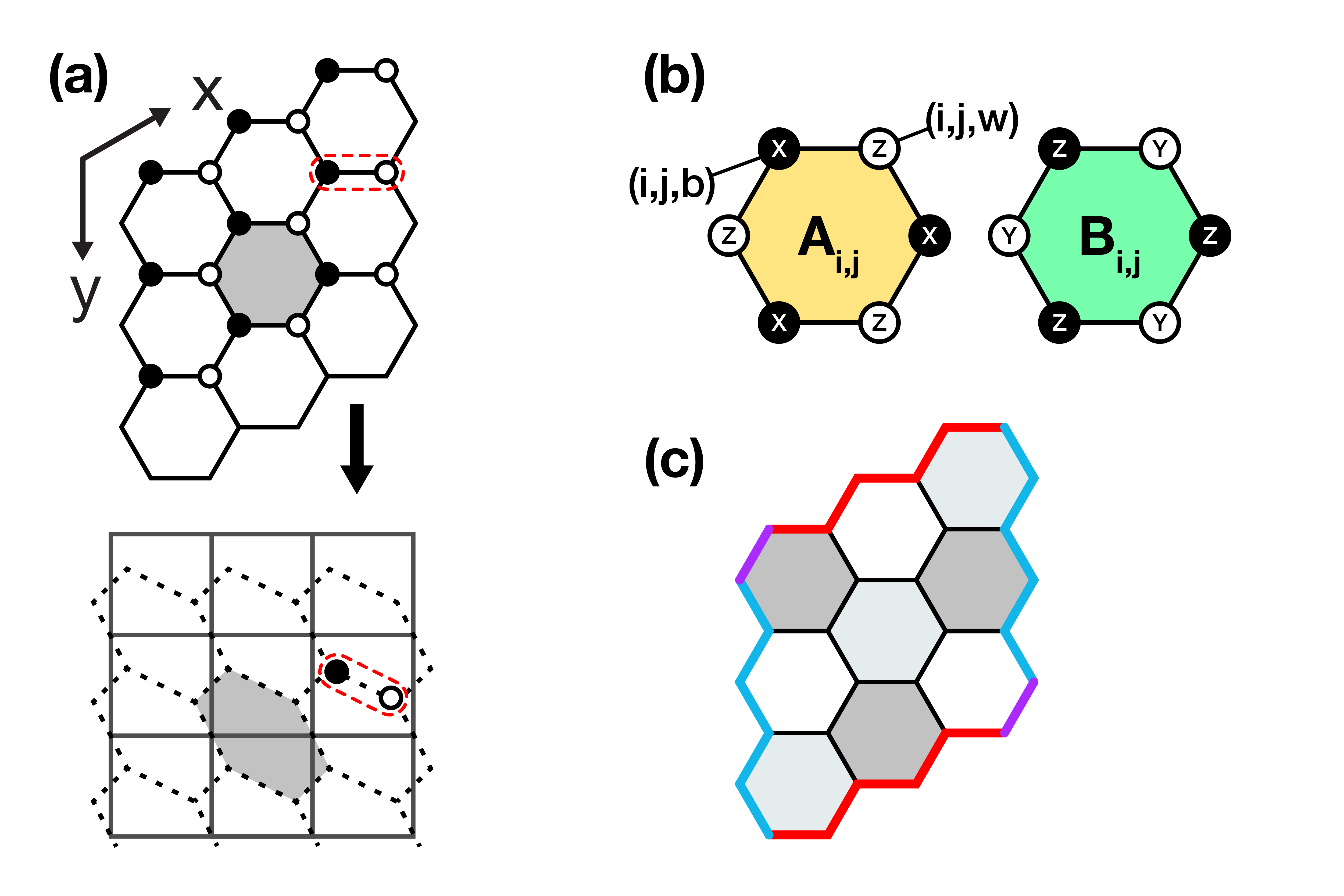}
    \caption{Stabilizers and layout of the XYZ color code. (a) A bipartion of the honeycomb lattice, and its mapping to the square lattice using a two-qubit unit cell (circled in red). The shaded hexagon illustrates how the lattice is deformed to fit a square geometry. Partitions are indicated by white and black circles, respectively. (b) Stabilizers $A/B_{i,j}$ are defined as having qubit $(i,j,b)$ at the top left corner. (c) Identification of the edges of a 3$\times$3 honeycomb lattice to get a torus. Edges of the same color (red, blue, and purple) are identified.}
    \label{fig:colorcode}
\end{figure}

In this section, we introduce the XYZ color code. We show that, if we restrict ourselves to a noise model that is infinitely biased towards one type of Pauli error, the only allowed logical errors of this code are fully described by the evolution of an elementary cellular automaton. These biased logical operators are fractal-like, and not string-like, with weights growing faster than the linear system size. We next describe an efficient global decoder, which finds the most probable logical error at infinite bias.

\subsection{The Model}
The XYZ color code is a stabilizer code defined on a three-colorable honeycomb lattice, where a single qubit is placed on each vertex of the lattice. The vertices are bicolored\cite{kubica2015}, black and white, such that no two vertices of the same color are touching, see Fig.~\ref{fig:colorcode} (a). The code space of the XYZ color code is the simultaneous +1 eigenspace of all of the stabilizers of the code. Its stabilizers generate a group, $\mathcal{S} \subset \mathcal{P}_N$; which is an Abelian subgroup of the Pauli group acting on $N$ qubits such that $-1 \not\in\mathcal{S}$. Up to phases, the Pauli group is generated by the standard Pauli operators $X_q$ and $Z_q$ acting on qubits indexed by vertices $q$. It is also helpful to define the group of Pauli-Z operators $\mathcal{P}^Z_N \subset \mathcal{P}_N$. This is the group of operators that can be generated from the product of Pauli-Z operators only.

The stabilizer group for the XYZ color code is generated by $A$- and $B$-type stabilizers, $A_{p}$ and $B_{p}$, shown in Fig.~\ref{fig:colorcode}(b).  The $A$-\textit{type stabilizers} act on each plaquette with Pauli $X$ and $Z$ operators on the black and white vertices, respectively, and the $B$-\textit{type stabilizers} act with $Z$ and $Y$, respectively, on the black and white vertices. We concentrate on the lattice with periodic boundary conditions. We therefore identify the boundaries of the three-colorable lattice as shown in Fig.~\ref{fig:colorcode}(c).

The XYZ color code is equivalent under an on-site unitary circuit to the conventional two-dimensional color code \cite{bombin2006, bombin2007} defined on a hexagonal lattice. As such, it inherits the generic properties and code parameters of the CSS color code. For instance, on the torus, the code has four logical qubits\cite{bombin2007}, that are acted on by string-like logical operators. The standard color code is a Calderbank-Shor-Steane (CSS)-type code, meaning that the stabilizer group can be generated by operators that are each products of either only Pauli-$Z$ operators, or only Pauli-$X$ operators. We obtain the XYZ color code by applying the unitary rotation $U = \prod_{q\in \textrm{white}} U_q$ to the CSS color code where $U_q$ acts on the standard Pauli operators as $U_qX_qU_q^\dagger =Z_q$ and $U_qZ_qU_q^\dagger = Y_q$. Indeed, several papers have recently demonstrated improvements in threshold using a similar local change of basis, see e.g. Refs.~\cite{bonilla2021,darmawan2021,Tuckett19, roffe2022, srivastava2021}. A similar code was also obtained~\cite{Teo2014} by applying a Hadamard rotation to alternate qubits. We expect that this code behaves equivalently to the XYZ color code under noise that is biased in an appropriate choice of basis.

We also introduce some terminology to describe error correction with the XYZ color code. We consider states that lie outside the codespace, $E|\psi\rangle$ where $|\psi\rangle$ is a code state and $E$ is a Pauli error. We say that a \textit{defect}, or an excitation, lies at a plaquette $p$ unless both $A_{p} E|\psi \rangle = (+1) E|\psi \rangle$ and $B_{p} E|\psi \rangle = (+1) E|\psi \rangle$. The \textit{syndrome} is the configuration of defects on the entire lattice.

In the sublattice picture, it is helpful to visualize the lattice on an $L \times H$ rectangular geometry, with $L$ and $H$ each a multiple of three to maintain three colorability on the periodic lattice. We use the two-qubit unit cell shown in Fig.~\ref{fig:colorcode}(a) to describe coordinates on the honeycomb lattice. There are therefore $N = 2LH$ qubits. Qubits are labeled $(i,j,b)$ or $(i,j,w)$, where $b$ and $w$ indicate the black and white sublattices, respectively. Stabilizers $A_{i,j}$ and $B_{i,j}$ are the stabilizers acting on the plaquette with top left qubit $(i,j,b)$.

\subsection{Noise Models}
\label{sec:noisemodels}

We consider several different noise models. In each noise model, we begin with a system in a logical state and take Pauli noise that acts independently on every qubit. Each Pauli error, $X$, $Y$, or $Z$, occurs as a Poisson process with rates $\gamma_X$, $\gamma_Y$, and $\gamma_Z$, respectively. If we wait for a fixed time, say, between calls to a global decoder, then there are probabilities $p_X, p_Y,$ and $p_Z$ of having an error on each qubit. It is clear that these probabilities grow with the error rates, and always lie between 0 and \(1/2\). We define $\gamma_{tot}$ and $p_{tot}$ as the sums of the rates and probabilities, respectively.

The main noise model we consider is \textit{infinitely biased noise}, with $\gamma_X = \gamma_Y = 0$, and $\gamma_Z > 0$. This error model is of particular interest when the code experiences a highly biased dephasing noise\cite{darmawan2021,geller2013}. However, an equivalent discussion holds for any choice of single Pauli operator, since the stabilizer group is symmetric under the interchange $X \rightarrow Z \rightarrow Y \rightarrow X$.

A related noise model is \textit{infinitely biased sublattice noise}, where we consider $Z$ noise only on the black sublattice. The reason we can do this is that at infinite bias, the sublattices effectively decouple. This is because Pauli-$Z$ operators acting on the black sublattice only excite the $A$-type stabilizers, whereas Pauli-$Z$ operators acting on the white sublattice only excite the $B$-type stabilizers. We use this noise model at some points for simplicity, but infinitely biased noise on the white sublattice is equivalent, up to spatial transformations.

In both cases of infinitely biased noise, we can describe the state of the system with a string of bits $s$. Let $\ket{\psi}$ be the starting logical state. After some time $t$ of accumulating errors, the state may be written
\begin{align}
    \ket{\psi(t)} = \prod_{q}Z_q^{s_q}\ket{\psi},
\end{align}
where $q=(i,j,b/w)$ is a qubit index and $s_q \in \{1,0\}$. This notation is helpful when defining the cellular automaton decoder.

Finally, we consider \textit{finitely biased noise}. In this setting, we allow $Y$ errors as well as $Z$ errors. Since $X=-iYZ$, the entire Pauli group $\mathcal{P}_N$ may be implemented by this error channel. The bias of the error channel is parameterized by the ratio $\zeta \equiv \gamma_Z/\gamma_Y$. For global decoding, we may also define $\zeta_p = p_Z/p_Y$.

\subsection{Logical Operators at Infinite Bias}
\label{sec:logicals}

\begin{figure}
    \centering
    \includegraphics[width=\columnwidth]{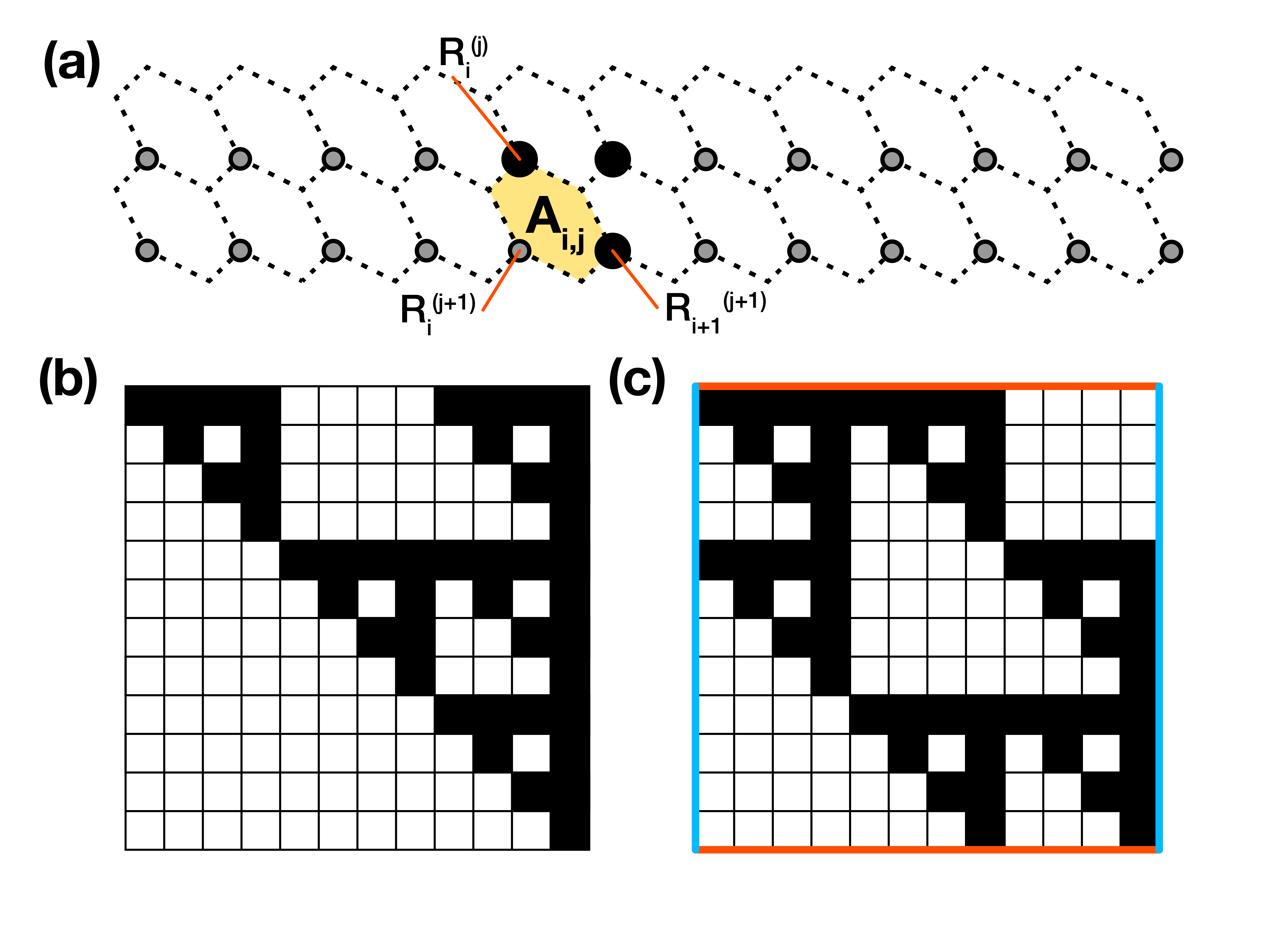}
    \caption{Evolutions of cellular automaton rule 102. (a) Two rows of the honeycomb lattice, showing how row stabilizer $A_{i,j}$ constrains $R^{(j)}$ and $R^{(j+1)}$. Larger circles indicate where a $Z$ operator is applied, i.e. where $R_i^{(j)}=1$. Only the black sublattice is shown. (b) Viewing the black sublattice as a square lattice, Sierpinski fractal created by evolving initial conditions 000000000001. Shaded squares represent $Z$ operators. (c) A configuration of rule 102 with periodic boundary conditions. The left and right edges, and the top and bottom edges, are identified. }
    \label{fig:CAevolutions}
\end{figure}

We refer to any operator that commutes with the stabilizer group, but has a nontrivial action on the code space, as a logical operator. In other words, the logical operators are the set $Z(\mathcal{S}) \backslash \mathcal{S}$, where $Z(\mathcal{S})$ is the centralizer of the stabilizer group.

We now consider what logical errors may occur under infinitely biased noise. These are the set of logical operators that are also members of $\mathcal{P}^Z_N$. We call such operators \textit{biased logical operators}. For simplicity, we restrict ourselves to infinitely biased single sublattice noise. We specify a logical operator $R$ using a series of bitstrings $R^{(j)}$. Each $R^{(j)}$ represents the action of the operator on row $j$, with bit $R^{(j)}_i$ being 1 if $R$ acts as $Z$ on site $(i,j,b)$, and 0 if it acts as the identity. Since stabilizer $A_{i,j}$ has only one $X$ operator on row $j$ and two on row $j+1$, $R^{(j)}$ must be completely determined by $R^{(j+1)}.$ In order to satisfy all stabilizers, we then require (Fig.~\ref{fig:CAevolutions}(a)):
\begin{align}\label{eq:nextrow}
    R^{(j)}_i=f(R^{(j+1)})_i\equiv R^{(j+1)}_i + R^{(j+1)}_{i+1},
\end{align}
where addition is performed mod 2. Additionally, due to periodic boundary conditions, indices of the bit string are only defined modulo $L$, where $L$ is the linear size of the system. The function $f:\mathbb{F}_2^L\rightarrow \mathbb{F}_2^L$ is called the update rule, or evolution rule. This is an example of a local update rule\cite{Yoshida13, yoshida2013, Devakul19}, or \textit{cellular automaton}: each bit in $R^{(j)}$ depends only on a local region in $R^{(j-1)}$. In particular, it is an example of an \textit{elementary cellular automaton}. 
Elementary cellular automata have been studied extensively and classified. In Wolfram's classification of elementary cellular automata, Eq.~\ref{eq:nextrow} is called rule 102 \cite{wolfram1983}. Starting from a bit string with only a single index equal to one, its evolution is a discrete form of the Sierpinski fractal, see Fig.~\ref{fig:CAevolutions}(b).

The Sierpinski fractal form of logical operators gives an intuitive explanation for the lack of string-like operators at infinite bias. In particular, the Sierpinski fractal has dimension $\log(3)/\log(2)$. If the system size is $L \times L $, then the weight of the fractal scales as $L^{\log(3)/\log(2)}$. A string-like logical operator, on the other hand, has weight scaling as $L$. We make this intuition rigorous in Appendix \ref{app:optimalsize}, by proving that there exists an infinite family of system sizes with no stringlike operators.

Due to the periodic boundary conditions, there is an additional constraint on which evolutions under rule 102 constitute logical operators. For an arbitrary row  $R^{(j)}$, we have $f^H\left(R^{(j)}\right)=R^{(j)}$, where $H$ is the height of the system. Fig.~\ref{fig:CAevolutions}(c) is an example of a configuration that respects periodic boundary conditions.

This condition cannot always be satisfied. For example,  $f^n\left(R^{(j)}\right)$ always has an even number of 1s, so if $R^{(j)}$ has an odd number of 1s, then it cannot satisfy any periodic boundary conditions. In fact, for an arbitrary system size, it is not obvious that any configurations can satisfy the periodicity constraint. In Appendix \ref{app:optimalsize}, however, we show that if the system has linear dimensions that are multiples of three, and are not both even, then there are always at least two independent logical operators in $\mathcal{P}^Z_N$ on each sublattice.
Indeed, lattices sizes that are multiples of three are required to satisfy the three-colorability constraint of the color code.
Furthermore, these logical operators are not equivalent to each other or the identity under multiplication by stabilizers, so they are distinct operators on the logical space. Additionally they are not string-like, with weight scaling as $LH/3$, i.e. linear in the total number of qubits.

Depending on the system size, there may also be additional logical operators. The valid configurations of elementary cellular automata with periodic boundary conditions are quite complex\cite{Yoshida13, Devakul19}, and, to our knowledge, an analytic formula for the number of such configurations for arbitrary system size has not been found. The number of valid periodic configurations has been calculated for certain system sizes, and it oscillates wildly with small changes in  size\cite{martin1984}. This can be seen as a classical analogue to the highly oscillatory behavior of the ground state degeneracy in some fracton models\cite{haah2011,haah2011energy,yoshida2013}.

Nevertheless, in Appendix \ref{app:optimalsize}, we show that for an infinite family of system sizes, there are exactly two independent biased logical operators. Unlike the typical case with CSS codes, each of these is the only representative of its equivalence class under multiplication by stabilizers.

Since these two operators have weight scaling as $N$, it means the code also has $\mathcal{O}(N)$ distance scaling at infinite bias. We also find an infinite \textit{one-parameter }family within this larger family, of system sizes with $L=3(2^n)$ and $H=3(2^n+1)$.

\subsection{Optimal Decoder}
\label{sec:exactdecoder}

We require a decoding algorithm to read out encoded logical information. Here we concentrate on decoding the effective fractal code in the limit of infinite bias. Earlier work has demonstrated decoding algorithms for fractal codes based on clustering~\cite{Bravyi2013}, and minimum-weight perfect matching~\cite{Nixon2021}. Here we propose an optimal decoder for the XYZ color code at infinite bias.

Our decoder is a maximum likelihood decoder, returning the lowest weight error in $\mathcal{P}^Z_N$ that causes a given syndrome. In Appendix \ref{app:proofthreshold}, we prove the following lemma, which shows that the decoder has a 50\% threshold:

\textbf{Lemma 1}: consider a family of error correction codes $\mathcal{C}$, parameterized by system size $N$. Let the error channel be i.i.d. and infinitely biased. Define $\mathcal{L}(N)$ to be the set of biased logical operators. If $|\mathcal{L}(N)|$ is constant with $N$, and if every error in $\mathcal{L}(N)$ has support polynomial in $N$, then a maximum likelihood decoder for $\mathcal{C}$ has a 50\% threshold.
\break

This lemma holds for any of the special system sizes mentioned in the previous section, where there exist a constant number of biased logical operators with weight scaling as $\mathcal{O}(N)$. Our decoder therefore has a 50\% threshold for these system sizes, the maximum value possible for i.i.d. biased noise.

Before describing the decoding algorithm itself, we briefly discuss notation. The decoder corrects errors on the black and white sublattices independently, so we can once again restrict to single sublattice noise. We again write an operator $E \in \mathcal{P}^Z_N$ as a string of bits $E^{(j)}_{i}$ that is 1 if $Z$ acts on site $(i,j,b)$, and 0 otherwise. We use this notation to write errors and correction operators.

We are also interested in the syndrome of an operator $E$, denoted $S^{(j)}_i(E)$, which is 1 if the $A$-type stabilizer $A_{i,j}$ is excited by $E$, and 0 otherwise. We neglect the $B$-type stabilizers as they are not excited by errors on the sublattice of interest.

Our decoding algorithm has 3 steps:
\begin{enumerate}
    \item Find an operator $C_1$ that moves all of the defects of the syndrome onto row $0$.
    \item Find an operator $C_2$ such that $S(C_2) = S(C_1 E)$.
    \item Return the most likely error $C$ such that $S(C) = S(C_1 C_2) = S(E)$. 
\end{enumerate}

We now explain how each of these steps may be computed efficiently.

\subsubsection{Moving Syndromes onto Row 0}
When we apply a $Z$ operator to site $(i,j,b)$, it creates excitations at $A_{i,j}$, $A_{i,j-1}$, and $A_{i-1,j-1}$. If there is already a defect on $(i,j,b)$, it moves the defect onto two sites in the row above. Starting from the bottom row, $S^{(H)}$, we then apply $Z$ to every site where the syndrome is 1. We repeat row by row until we get to row 0. This defines $C_1$. The remaining syndrome $S(C_1E)$, is 0 on all rows except for $S^{(0)}$.

\subsubsection{Finding the operator $C_2$}
We now want to invert the excitation map; in other words, find an operator such that its syndrome is identical to that of $C_1E$. In general, this problem may be difficult, but because we have moved all of the syndromes onto one line, this takes a particularly simple form. To see this, note that $S^{(j)}(E)=0$ implies that
\begin{align}\label{eq:nextrowcorrection}
    E^{(j)}_i=E^{(j+1)}_i+E^{(j+1)}_{i+1}.
\end{align}
This is again cellular automaton rule 102, given in Eq.~\ref{eq:nextrow}. This tells us that $E^{(j)} = f\left(E^{(j+1)}\right)$, where $f$ is the evolution rule generating Eq.~\ref{eq:nextrowcorrection}. Since $S(C_1E)$ vanishes on all rows except 0, this then means that $C_1E^{(1)}=f^H\left(C_1E^{(0)}\right)$. The operator $C_1E^{(0)}$ now satisfies the following equation:
\begin{align}\label{eq:c2condition}
    S^{(0)}_i &= C_1E^{(0)}_i + C_1E^{(1)}_i + C_1E^{(1)}_{i+1}\\\nonumber
    &=C_1E^{(0)}_i + f^H\left(C_1E^{(0)}_i\right)+f^H\left(C_1E^{(0)}_{i+1}\right).
\end{align}
By Eq.~\ref{eq:nextrowcorrection}, $f$ is linear. We may therefore solve Eq.~\ref{eq:c2condition} for $C_1E^{(0)}$ efficiently using Gaussian elimination. Note that the solution given is not generally unique or equal to $C_1E^{(0)}$; this degeneracy reflects the fact that there are logical operators and possibly products of stabilizers that can be implemented by infinitely biased noise. However, at this stage we only care about finding a \textit{potential} correction, and not finding the most likely correction.

We label our solution $C_2^{(0)}$. To find the rest of $C_2$, we simply use the evolution rule in Eq.~\ref{eq:nextrowcorrection}.

\subsubsection{Returning the most likely correction}
We now look for alternative corrections that may be more likely. To do this, we simply enumerate over all possible logical operators. Using the results from Appendix~\ref{app:optimalsize} and choosing the right system size, we only need to consider two independent logical operators in the infinite bias case. Taking products, as well as the identity, means there are four possibilities. These four potential corrections all represent inequivalent operators on the logical space. The most likely error therefore represents the most likely logical state before errors occurred. We then apply each logical operator $R$ to $C_1C_2$, and return the $RC_1C_2$ with the lowest weight.
\break

The runtime of this decoding algorithm is polynomial in the system size. The first step has a runtime of $\mathcal{O}(LH)$. In the second step, the runtime for the Gaussian elimination is $\mathcal{O}(L^2)$, and the time to evolve the error $C_1E^{(0)}$ onto all other rows is $\mathcal{O}(LH)$. In the final step, the runtime is $\mathcal{O}\left(N_L LH\right)$, where $N_L$ is the number of biased logical operators. Since in the previous section we mentioned a one-parameter family of system sizes with two independent biased logical operators, $N_L$ can be made to be constant, see Appendix~\ref{app:optimalsize}.

\section{Cellular Automaton Decoder}
\label{sec:cadecoder}

A local decoder is a local quantum channel which, when applied on top of an error channel, reduces the logical error rate. A local decoder cannot share quantum or classical information across arbitrarily large distances. In particular, if the decoding operation applies a correction operator $C_q$ to qubit $q$, then $C_q$ cannot depend on any stabilizer measurement outside of some local neighborhood of $q$, independent of system size.

In this section, we present a local decoder for infinitely biased noise in the XYZ color code. First, however, we briefly review the Newman Moore model~\cite{newman1999}, a classical spin Hamiltonian that exhibits an energy barrier between ground states that is logarithmic in linear system size. We then derive a \textit{probabilistic} local decoder, under which the ensemble of states of the XYZ color code converges to a thermal distribution for the Newman Moore model. We use this relationship to derive a duality between the error rate $\gamma_Z$ of the quantum memory and the inverse temperature $\beta$ of the Newman-Moore model.

\subsection{The Newman-Moore Model}

\begin{figure}
    \centering
    \includegraphics[width=\columnwidth]{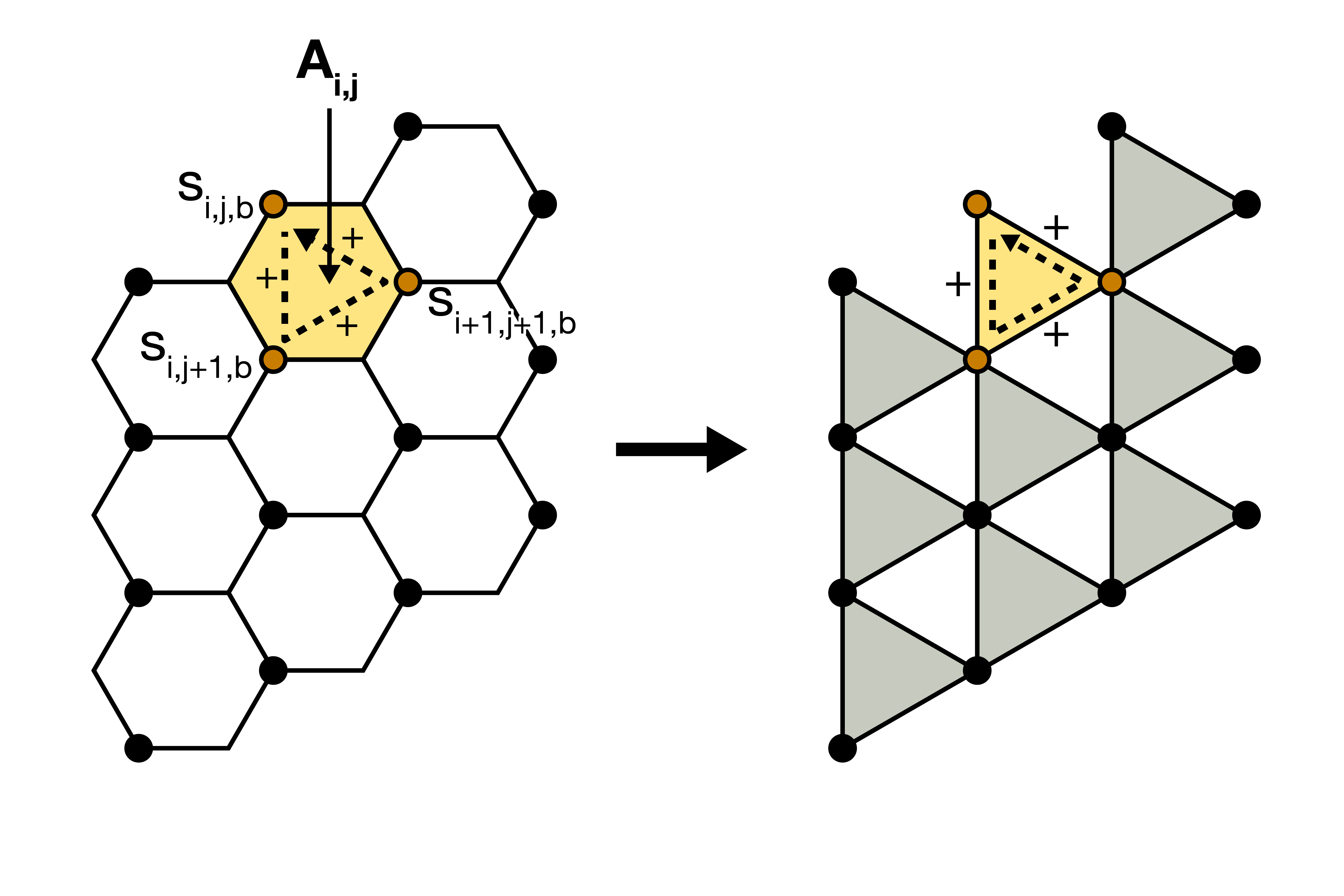}
    \caption{Relationship between the $A$-type stabilizers at infinite bias (left) and the Newman-Moore model (right). Arrows indicate terms that are added together modulo 2. The $B$-type stabilizers have a similar relation to a flipped Newman-Moore model.}
    \label{fig:duality}
\end{figure}

We find a simple relationship between the Newman-Moore model and the XYZ color code at infinite bias, and we use this relationship to design a cellular automaton decoder for the XYZ color code at infinite bias that mimics the Newman-Moore model. The Newman-Moore model\cite{newman1999} is a classical statistical mechanical model with fragile glassy dynamics\cite{Chamon2005,castelnovo2012, Bravyi2013, brown2016} and no randomness, see also Refs.~\cite{Yoshida13, yoshida2013, Devakul19}. Its ground states have been shown to be separated by an energy barrier that scales logarithmically with system size\cite{newman1999, garrahan2000}. This energy barrier can protect logical information encoded in the ground states of the system in a thermal environment. This is because it is energetically unfavorable for the environment to introduce an error that will cause error correction to fail.

The degrees of freedom of the Newman-More model are spins $\sigma_i = 0,1$, defined on a triangular lattice. The Hamiltonian is a sum of local three-body terms, where each term is a sum of the spins $\sigma_j = 0,\,1$ on a right-facing triangle, see Fig.~\ref{fig:duality}:
\begin{align}\label{eq:originalnewman}
    H_{NM} = \sum _ {i,j,k \textrm{ in } \triangleright} \left(\sigma_i+\sigma_j+\sigma_k \mod 2\right).
\end{align}

Let us now look at the relationship between the Newman-Moore model and the XYZ color code. We consider the Hamiltonian
\begin{align}
    H_{AB} = -\sum_{i,j}A_{i,j} + B_{i,j}, \label{eq:HAB}
\end{align}
whose ground states are the code space of the XYZ color code. If we take the error channel to be infinitely biased single sublattice noise, then we need only consider excitations of the $A_{i,j}$ terms.

The $A_{i,j}$ terms of $H_{AB}$ are only excited by Pauli-$Z$ errors on the three sites where their support is a Pauli-$X$ term. Using the bit string notation $s_{i,j,b}$ defined in Sec.~\ref{sec:noisemodels}, Eq.~\ref{eq:HAB} becomes:
\begin{align}
    H_{AB}\ket{\psi}=\sum_{i,j}\left(s_{i,j,b}+s_{i+1,j,b}+s_{i+1,j+1,b} \mod 2\right)\ket{\psi}.
\end{align}
such that the energy eigenvalues of $H_{AB}$ are those of $H_{NM}$ given in Eq.~\ref{eq:originalnewman}, where we replace the labels for the spin variables.

The mapping we have described is illustrated in Fig.~\ref{fig:duality}. The three-body term is again composed of spins at the vertices of right facing triangles, so this Hamiltonian is exactly the Newman-Moore model. If we instead consider the Hamiltonian with both stabilizers, we get two disjoint copies of the Newman-Moore model.

More generally, the duality in the next section uses a slightly different, but morally equivalent, stabilizer Hamiltonian,
\begin{align}\label{eq:symmetricnewman}
    H = \sum_{i,j}A_{i,j} + B_{i,j}+ A_{i,j}B_{i,j}.
\end{align}
This Hamiltonian has the advantage of being symmetric under exchange of $X$, $Y$, and $Z$ errors, so it performs equally against noise that is biased to introduce Pauli-X, Pauli-Y or Pauli-Z errors.

To summarize, the code space of the XYZ color code at infinite bias is dual to the ground space of a classical Hamiltonian, the Newman-Moore model. We note that such a duality is a standard property of bias-tailored codes\cite{bonilla2021, Tuckett2018, Tuckett19}, but the features of the target Hamiltonian depend critically on the choice of initial code. For example, in the more common bias-tailored surface codes, the emergent classical model is a product of 1D Ising Hamiltonians. These Ising Hamiltonians lead to a repetition code structure\cite{Tuckett19} that allows for a 50\% threshold, just as in the XYZ color code. However, the XYZ color code has an additional feature: its classical Hamiltonian hosts a logarithmically growing energy barrier, while the energy barrier of the 1D Ising model is constant. For this reason, we expect that only the XYZ color code will host a local decoder.

\subsection{Probabilistic Cellular Automata}

We propose a cellular automaton decoder that simulates the thermal dynamics of the Newman-Moore model at infinite bias. We achieve this using a local circuit of stabilizer measurements and Pauli-Z rotations, where the flips happen according to some probability distribution dictated by the local measurements. This circuit is an example of a \textit{stochastic cellular automaton}. 

At the most general level, a stochastic, or probabilistic cellular automaton is a sequence of states $s^{(T_0)},s^{(T_1)},...$ following a local update rule, but with the addition of randomness. Specifically, the state $s^{(T_{i})}$ is no longer simply a deterministic function of $s^{(T_{i-1})}$, but may depend on some random variables as well.

We define a stochastic cellular automata to be a \textit{continuous time Markov chain} with a few conditions on locality. Namely, the states are bit arrays, the only nonzero transition rates are between bit arrays that differ only in a local neighborhood of a point, and the transition rates themselves are only functions of the state in the same neighborhood. We now make each of these statements precise.

A continuous time Markov chain\cite{norris1997} is defined by a set of states $\{s^{(1)},s^{(2)},...,s^{(n)}\}$, and an $n\times n$ transition matrix $\Gamma$. State $s^{(i)}$ transforms to state $s^{(j)}$ following a Poisson process with rate $\Gamma_{ij}$. The first condition for a continuous time Markov chain to be a stochastic cellular automaton can be stated as follows. If $\Gamma_{ij}$ is not zero, then there must exist an index $q$ such that
\begin{align}
    s^{(i)}_k = s^{(j)}_k, k \notin B_R(q),
\end{align}
where $B_R(q)$ is a ball of radius $R$ around $q$ and $R$ does not depend on system size. The second condition is that the transition rates $\Gamma_{ij}$ are a local function of the states,
\begin{align}
    \Gamma_{ij} = f\left(s^{(i)}_k,s^{(j)}_k | k \in B_R(q)\right),
\end{align}
with $R$ and $q$ chosen as before. In other words, if the state changes near some point, the rate of that transition must only depend on the state restricted to a neighborhood of that point.

A \textit{local decoder} is a local cellular automaton applied to some noisy code state $E\ket{\psi}$ where $E$ is some Pauli error, and $\ket{\psi}$ is a logical state. At infinite bias, we again use the notation in Sec.~\ref{sec:noisemodels} to represent $E\ket{\psi}$ as a bit string $s$. We use a cellular automaton that only applies $Z$ to one qubit at a time. That is, the only process that occurs with nonzero rate is $\ket{\psi} \rightarrow Z_q\ket{\psi}$. As a shorthand, we define the rate for this process as 
\begin{align}
    \gamma_q(s) \equiv \Gamma_{s,s \oplus \hat{q}},
\end{align}
where $\hat{q}$ is the string with 1 at index $q$ and 0 elsewhere. This cellular automaton must run without destroying logical information. Therefore, it can only measure stabilizers. We then require that $\gamma_q$ is a function only of the stabilizers in a neighborhood of $q$, and not any more general function of $S$.

\subsection{Cellular Automaton Decoder}
We now define a probabilistic cellular automaton rule that maps onto thermal fluctuations of the Newman-Moore model. At infinite bias, the total transition rate $G_q$, or the total rate of a $Z$ flip occurring on qubit $q$, is given by the sum of the cellular automaton transition rate and the error rate
\begin{align}
    G_q = \gamma_q + \gamma_Z.
\end{align}

Let $\omega$ be the change in energy of the Hamiltonian in Eq.~\ref{eq:symmetricnewman} upon applying $Z_q$. Since $H$ is local, $\omega$ is only a function of local stabilizers in a neighborhood of $q$, so $\gamma_q$, and therefore $G_q$, may depend on $\omega$. We then require that $G_q$ satisfies the \textit{detailed balance condition}:
\begin{align}\label{eq:detailedbalance}
    G_q(\omega) = e^{\beta\omega}G_q(-\omega),
\end{align}
for some $\beta > 0$. It is easily shown that if the transition rates of any Markov chain meet this condition, then the thermal distribution
\begin{align}
    P(s)=\frac{1}{\mathcal{Z}}e^{-\beta H(s)},
\end{align}
is a fixed point distribution of the chain. Here $\mathcal{Z}$ is the partition function of $H$ at inverse temperature $\beta$.

There is still a great degree of freedom in satisfying Eq.~\ref{eq:detailedbalance}. First, there is the choice of $G_q(\omega)$. A simple solution to Eq.~\ref{eq:detailedbalance} is given by\cite{brown2016,kossakowski1977}
\begin{align}\label{eq:tottransitionrate}
    G_q(\omega) = \frac{\omega}{1-e^{-\beta\omega}}.
\end{align}
With this solution, the transition rates of the cellular automaton rule are
\begin{align}\label{eq:transitionrate}
    \gamma_q(\omega) = \frac{\omega}{1-e^{-\beta\omega}}-\gamma_Z.
\end{align}

Next there is the choice of $\beta$. Any $\beta$ may be used as long as Eq.~\ref{eq:transitionrate} is always greater than 0. It makes sense to use the largest possible value of $\beta$, since this should increase the lifetime of any ground state. We then choose $\beta$ such that the smallest $\gamma_q(\omega)$ is 0. From Eq.~\ref{eq:symmetricnewman}, $\omega$ takes values in the range $[-6,+6]$. Therefore, we require
\begin{align}
    \frac{-6}{1-e^{6\beta}}=\gamma_Z,
\end{align}

or
\begin{align}\label{eq:gammabetadual}
    \beta = \frac{1}{6}\ln\left(\frac{6+\gamma_Z}{\gamma_Z}\right).
\end{align}

Under the local decoder and infinitely biased noise, the XYZ color code is then dual to a Newman-Moore Hamiltonian, see Eq.~\ref{eq:symmetricnewman}, with an inverse temperature that diverges logarithmically as the error rate vanishes.

\section{Demonstration of Partial Self-Correction}
\label{sec:infiniteresults}

\begin{figure}
    \centering
    \includegraphics[width=\columnwidth]{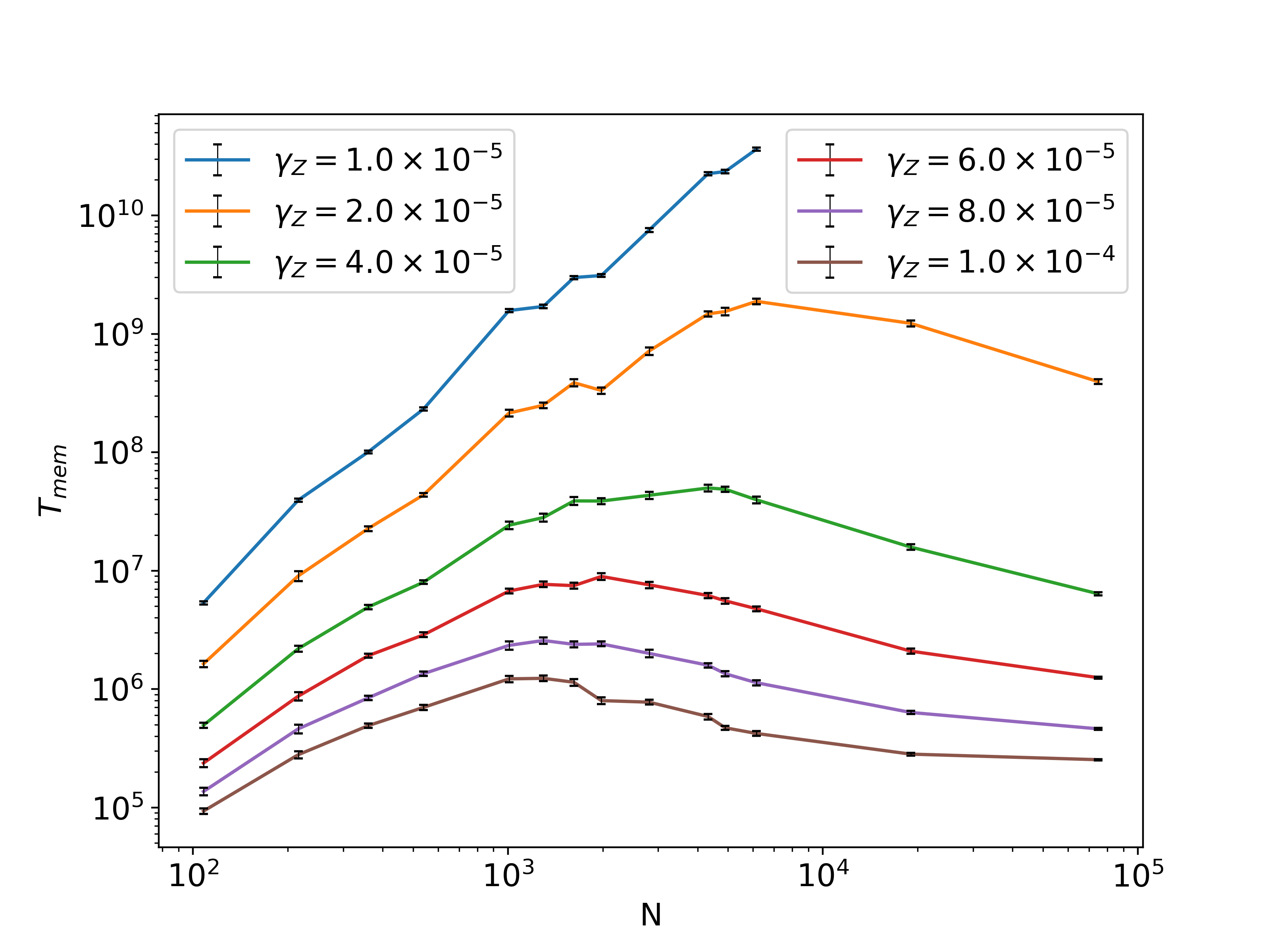}
    \caption{Memory time vs system size. The growth, and then decline, of memory time with system size is clearly visible.}
    \label{fig:memvssystemsize}
\end{figure}

We look to determine the performance of our cellular automaton decoder. In practice, this decoder will need to be supported by a global decoding system where the cellular automata operate at a high frequency, and the global decoder is called at a slower rate that is determined by communication speed and bandwidth limitations. Here, however, we focus on the performance of the cellular automaton decoder alone. We evaluate its performance by asking for how long the XYZ color code can maintain its logical information using the cellular automaton decoder for various noise parameters and system sizes.

In this section, under an infinitely biased noise model, we demonstrate that the cellular automaton decoder we have proposed can reproduce the behavior of a partially self correcting memory. Partial self correction is a phenomenon first found in fracton codes\cite{haah2011energy, haah2013}, characterized by memory time growing quickly with system size up to some critical linear system size $L_C$, which diverges with $\beta$. After reaching this size, memory time then falls off slowly.

This scaling lies in an intermediate regime towards true self correction, where quantum information is stored in a bistable phase at finite temperature. Self-correcting memories, such as the four-dimensional toric code\cite{alicki2010} have a memory time that grows exponentially with system size below a critical temperature. However, partial self-correcting memories are more stable than memories such as the two-dimensional toric code\cite{alicki2010}, that have only a constant energy barrier between ground states.

There are two scaling laws for memory time that we use as signatures to demonstrate partial self correction\cite{haah2013,brown2016}. First, the maximum memory time $T_{max}$, or the memory time at $L=L_C$, scales exponentially in $\beta^2$:
\begin{align}\label{eq:membetascaling}
    T_{max} \sim \exp(a\beta^2),
\end{align}
where $a$ is a positive constant. Second, at system sizes far below $L_C$, the memory time grows as a power law in $L$, where the exponent is proportional to $\beta$:
\begin{align}\label{eq:memLscaling}
    T_{mem} \sim L^{C\beta},
\end{align}
where $C$ is another positive constant. We numerically identify these signatures of partial self correction using our cellular automaton decoder at infinite bias.

\subsection{Methodology}

\begin{figure}[t!]
    \centering
    \includegraphics[width=\columnwidth]{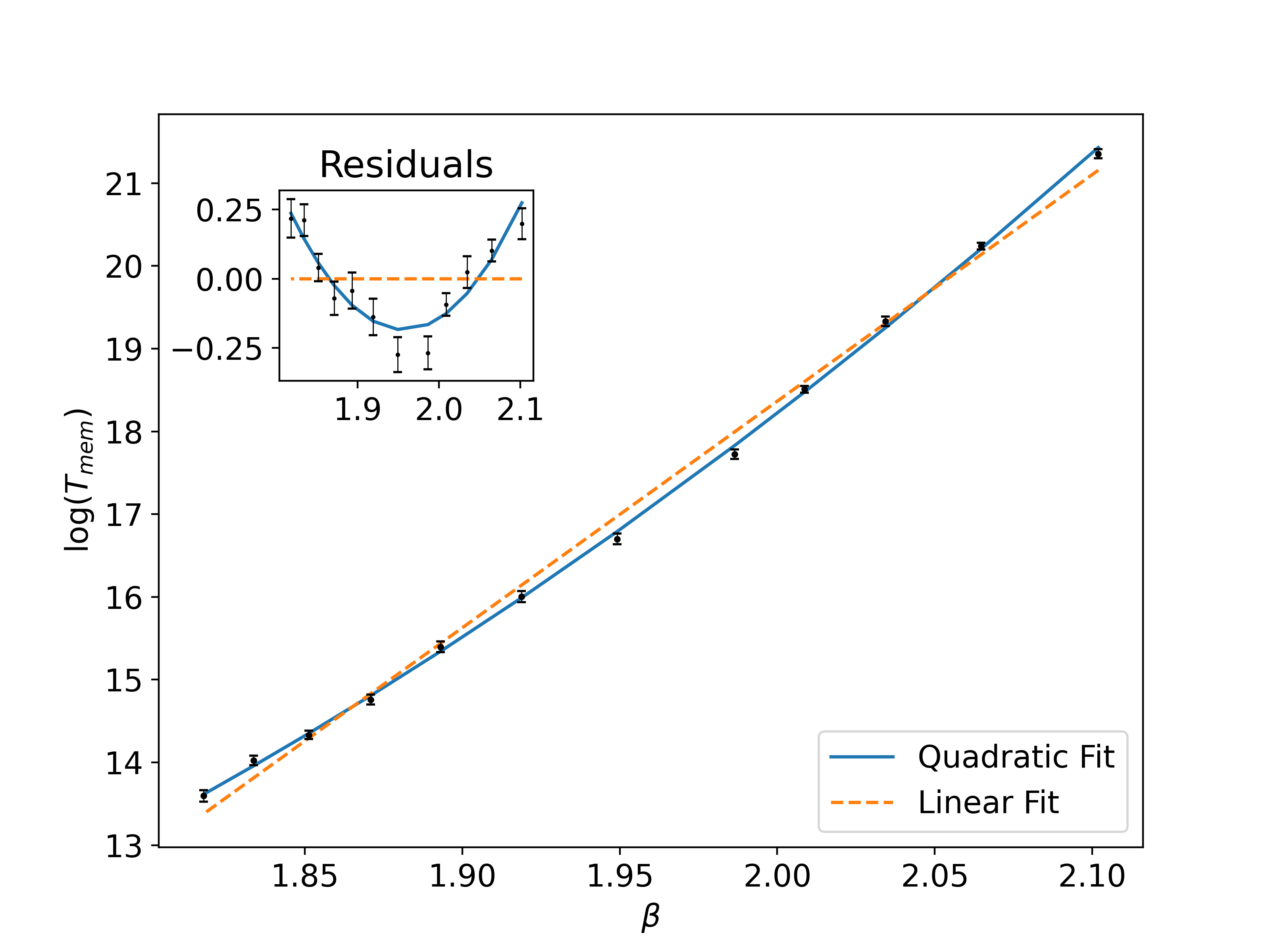}
    \caption{Logarithm of maximum memory time vs temperature. The best quadratic and linear fits are shown. The inset shows the residuals from the best linear fit.}
    \label{fig:quadfit}
\end{figure}

We simulate the dynamics of the cellular automaton decoder and noise channel with the Bortz-Kalos-Lebowitz (BKL) algorithm\cite{bortz1975}. This algorithm is well-suited to simulating continuous-time processes, such as the Poisson processes defined in Sec. \ref{sec:cadecoder}. Our simulation code is provided in a github repository in Ref.~\cite{sanmiguel2023}.

To evaluate the memory time, we choose an interval $\Delta t$, and run the global decoding algorithm at intervals approximately \( \Delta t\) apart during the BKL algorithm, to determine if the logical information can still be recovered. We do not apply the correction resulting from the global decoder, but instead check if this correction would create a logical error. We repeat this sequence until a logical error is created, and the total time is recorded as the memory time $T_{mem}$.

The interval $\Delta t$ between exact decoding sequences can be made larger to speed up the simulation. In order not to severely impact precision, it is chosen to be a fraction no larger than $10^{-3}$ of the total memory time. Due to the nature of the BKL algorithm, the time between exact decoding steps is not exactly $\Delta t$. Instead, an exact decoding step is performed after the first spin flip that occurs at a time greater than $\Delta t$ after the last exact decoding step. Typically, this is still very close to $\Delta t$.

In order for the exact decoder to be efficient, we use system sizes with minimal degeneracy of logical operators, as shown in Appendix \ref{app:optimalsize}. We computed the number of independent biased logical operators before running the simulations, and checked in each case that there are four per sublattice.

We calculate the memory time by evaluating the half life, or the median memory time over many samples collected for each data point.
Each data point is obtained using between
100 and 2000 samples, depending on system size. For consistency, we have only chosen system sizes of the form \begin{align}
L\times H = 3n \times 3(n+1)
\end{align}
for certain values of $n$. The set of system widths $L$ we use is:
\begin{align}
    L=6,9,12,15,21,24,27,30,36,45,48,54,96,192.
\end{align}

\subsection{Partial Self Correction}

Our numerical data that demonstrates partial self correction for the infinitely biased noise model is shown in Fig.~\ref{fig:memvssystemsize}. We now show that this scaling satisfies Eqs. \ref{eq:membetascaling} and \ref{eq:memLscaling}. This is consistent with the behavior of partially self-correcting memories.

The first signature of partial self-correction is the scaling of maximum memory time $T_{max}$ with temperature in Eq.~\ref{eq:membetascaling}. We find
\begin{align}\label{eq:quadfit}
    T_{max} &= \exp \left(a\beta^2+b\beta+c\right), \\
    a &= 21.33 \pm 2.87, 
\nonumber\\ b &= -56.10 \pm 11.23, \nonumber \\
    \nonumber
    c &= 45.12 \pm 10.98.
\end{align}

A plot of the maximum memory time and best fit is shown in Fig.~\ref{fig:quadfit}. Here, units are used where the coefficient of each term in the Newman-Moore model is 1.

The quadratic dependence may be contrasted with the Arrhenius-law scaling found in a code with no self-correction~\cite{brown2016}. In that case, the maximum memory time simply scales as $\exp(b\beta)$, with $b$ on the order of the energy gap. A best linear fit of the exponent is also shown in Fig.~\ref{fig:quadfit}. For that fit
\begin{align}
    T_{max} &= \exp \left(b\beta+c\right), \\
    b &= 27.36 \pm 0.51, 
\nonumber \\
    c &= -36.3 \pm 1.0.
    \nonumber
\end{align}

We find that $b$ is much larger than the gap, which is 6, see Eq.~\ref{eq:symmetricnewman}.

The next signature of partial self-correction comes from the power law scaling of memory time with system size at small sizes, in Eq.~\ref{eq:memLscaling}. This scaling is clearest at error rates (and thus temperatures) much lower than those in Figs. \ref{fig:memvssystemsize} and \ref{fig:quadfit}. Because the maximum memory time $T_{max}$ occurs at larger and larger system sizes as temperature decreases, this means that lower temperatures will have a larger region where the memory time grows exponentially.

\begin{figure}[t!]
    \centering
    \includegraphics[width=\columnwidth]{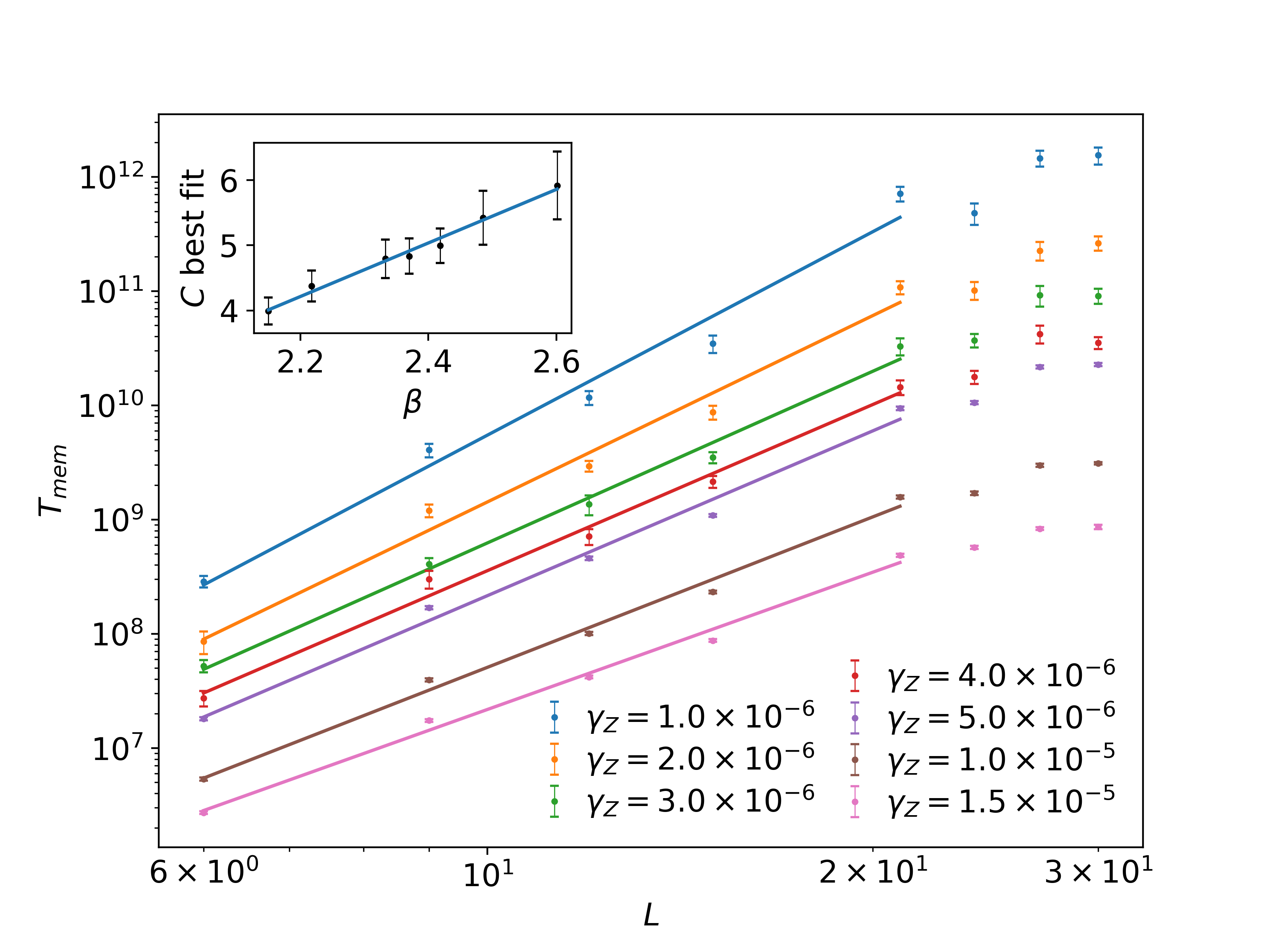}
    \caption{Memory time vs system size for small sizes and low error rate, with exponential fits. The inset shows the variation of the exponent with $\beta$.}
    \label{fig:linfit}
\end{figure}

The growth of memory time with linear system size for small sizes is shown in Fig.~\ref{fig:linfit}. We use system sizes ranging from $6 \times 9$ to $24 \times 27$, where $T_{mem}$ is far below $T_{max}$ at the relevant temperatures. We find that at small system sizes,
\begin{align}\label{eq:linfit}
     T_{mem} &\sim L^{C\beta+D}, \\
    C &= 4.10 \pm .22, \nonumber \\
    D &= -4.81 \pm .51. \nonumber
\end{align}

In both Eqs. \ref{eq:quadfit} and \ref{eq:linfit}, we have found similar scaling to fracton codes in three dimensions. In Ref.~\cite{haah2013}, such scaling has been proven analytically as a lower bound for Haah's cubic code, as well as other topological codes with a logarithmic energy barrier between code states. The proof involves the use of a general-purpose renormalization group decoder. Here, the Newman-Moore model also exhibits a logarithmic energy barrier, and we use an exact decoder, which must perform at least as well as the renormalization-group decoder. Therefore, it is encouraging that the XYZ color code reproduces these same signatures at infinite bias.

Finally, we note the small oscillations of the curves in Figs. \ref{fig:memvssystemsize} and \ref{fig:linfit}. These oscillations cannot be fully explained by random error, as they correlate between different curves and are greater than the sample error. Similar patterns are also seen in fracton codes \cite{haah2013, brown2016}. Regardless, partial self correction occurs at much larger time scales, and is therefore robust to these variations.

\section{Finite Bias}
\label{sec:finitebias}
We now consider the more realistic case of finite bias, where Pauli-Z errors are predominant, but are not the only type of error. We look to determine if the behavior we witnessed at infinite bias is robust when $\zeta = \gamma_Z/\gamma_Y$ is large, but not infinite.

To summarize our results, we find that finite bias inhibits the partially self-correcting behavior we observed where other types of error occur at a rate that is much slower than the memory time of the system. Indeed, this is consistent with the observations of other recent work~\cite{higgott2022}, where it is shown that the logical failure rate scaling is highly sensitive to a small amount of finite bias. However, we find that there are some modest  improvements to the memory time that persist at physically relevant biases. 

First, we find that our decoder demonstrates a threshold that improves with bias. We also show that the local cellular automaton decoder, which we stress is designed to correct for infinitely biased noise, still leads to an increase in memory time for low enough bias, although we do not observe that this improvement grows significantly with system size.

Since partial self correction is not a property of the thermodynamic limit, we do not expect the results of this section to define any sharp phase transitions. For example, it is always possible to find a bias high enough such that partial self correction occurs at some inverse temperature $\beta$. To see this, note that the median time for one $Y$ error to occur is given by the half-life of the exponential distribution,
\begin{align}\label{eq:TY}
    T_Y = \frac{\zeta\ln(2)}{N\gamma_Z},
\end{align}
where $N$ is the number of qubits. If $T_Y$ is much larger than the infinitely biased memory time, then most simulations will fail from $Z$ errors before even a single $Y$ error occurs. However, given the scaling of $T_{max}$ in Eq.~\ref{eq:quadfit}, this requires that $\zeta$ must grow at least as fast as $\exp(\beta^2)$. This is only a lower bound, since the critical system size $L_C$ increases with $\beta$ as well, and therefore this scaling may need to grow even faster to maintain an effective infinite bias limit.

This exponential scaling may be implausible in a physical architecture, so we instead study a constant range of biases, with $\zeta$ between 10 and 100. Such values are in line with some experimental qubit implementations\cite{darmawan2021,kjaergaard2020}. At these values, $T_Y$ is much smaller than $T_{max}$ for all of the inverse temperatures we study, and thus $Y$ errors can have a major effect on memory times.

In this regime, several difficulties appear. First, the exact decoder of Sec.~\ref{sec:exactdecoder} no longer provably finds a valid correction. In fact, it can be checked that a single $X$ or $Y$ error causes this decoder to fail. A second problem is that the exact duality with the Newman-Moore model no longer holds. With finite bias, all of the terms in Eqs. \ref{eq:originalnewman} and \ref{eq:symmetricnewman} become 6-body terms. Both of these issues are related to the fact that a $Y$ error creates a set of defects that cannot be created locally with $Z$ errors.

Most importantly, the full set of logical operators may be implemented by the error channel, including string-like logical operators. Therefore, the energy barrier becomes a constant for any finite bias, and the bounds from Ref.~\cite{haah2013} that define partial self-correction no longer hold.

\subsection{Methodology}
To simulate the XYZ color code at finite bias, we use the same cellular automaton decoder as in Sec.~\ref{sec:cadecoder}, along with the BKL algorithm. Our finite bias code is again included in Ref.~\cite{sanmiguel2023}. The total transition rate for $Z$ flips on qubit $q$, $G_q^{Z}$, is still given by Eq.~\ref{eq:tottransitionrate}. However, at finite bias, we add $Y$ errors to our rate equation as well, with a rate given by
\begin{align}\label{eq:yrate}
G^{Y}_q = \frac{1}{\zeta}\gamma_Z.
\end{align}
Pauli-$Y$ errors break the duality to the Newman-Moore model, as discussed previously. They also break detailed balance, since the rate of $Y$ flips in Eq.~\ref{eq:yrate} is not dependent on the energy of Eq.~\ref{eq:symmetricnewman}. We could restore detailed balance by introducing a cellular automaton rule that makes $Y$ flips as well as $Z$ flips, for example
\begin{align}
    \gamma^{Y}_q(\omega) = \frac{1}{\zeta}\frac{\omega}{1-e^{-\beta\omega}} -\frac{1}{\zeta}\gamma_Z.
\end{align}
However, there is little advantage to doing so. Since the Hamiltonian in Eq.~\ref{eq:symmetricnewman} no longer has a logarithmic energy barrier, even very small thermal fluctuations lead to logical errors. In our numerics, we were unable to find any benefits to such a strategy. For simplicity, we use the cellular automaton to make only $Z$-type corrections.

With our dynamics now specified, the next step is to find a decoder to evaluate success or failure of memory. For this, we use the renormalization-group decoder due to Bravyi and Haah\cite{haah2013}. This decoder, which works by identifying correctable clusters of errors, is proven to have a threshold for any translationally invariant topological code with independent and identically distributed single-qubit errors. Since the decoder does not take any information about the parameters of the local error channel, we use the same decoding procedure for all values of the bias.

\begin{figure}
    \centering
    \includegraphics[width=\columnwidth]{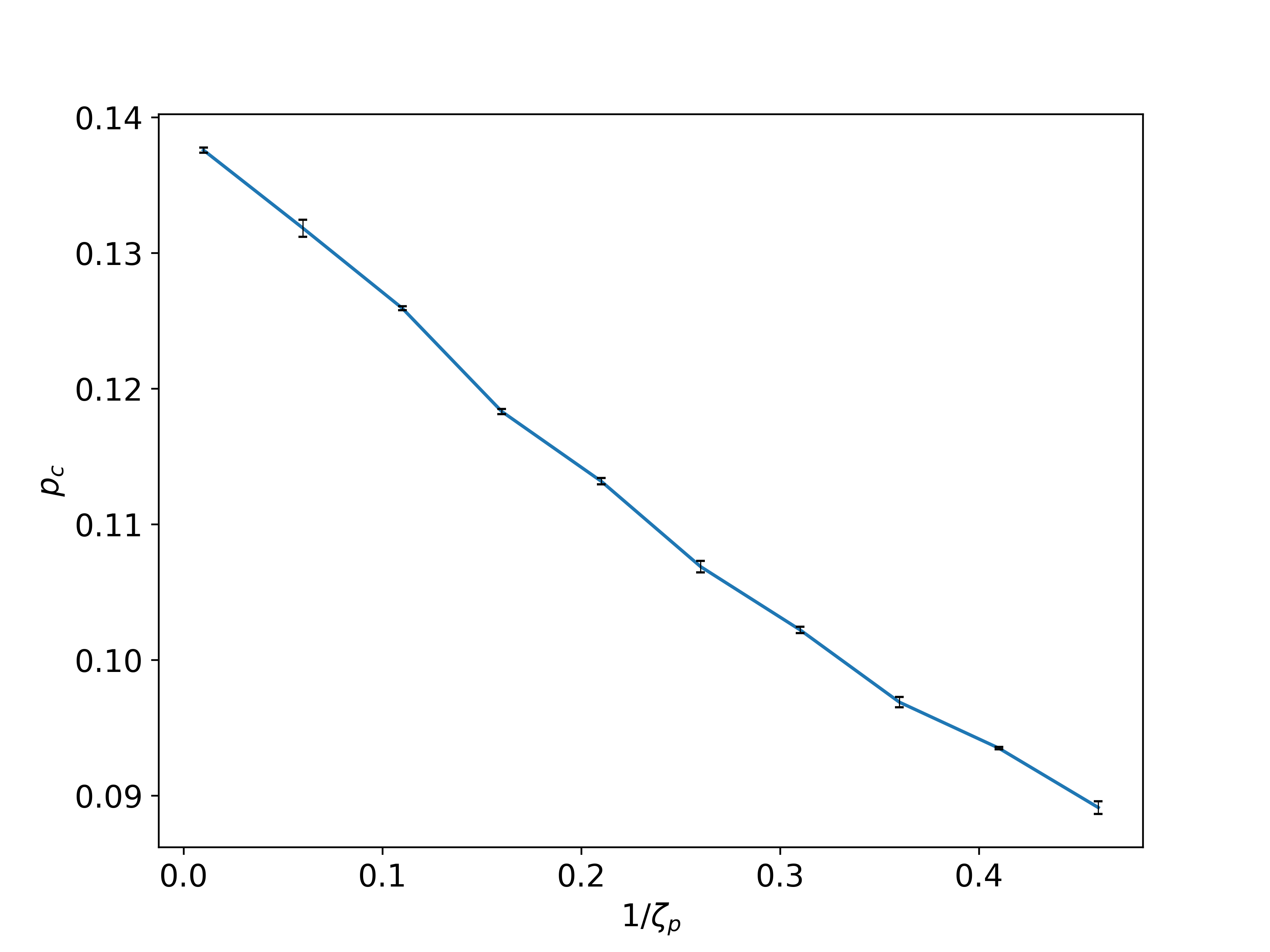}
    \caption{Threshold versus bias of the RG decoder. Threshold is calculated using systems with widths $L$=24, 48, and 96. The total error $p_{tot}$ for each width is varied between .05 and .15, and for each value, 10000 samples are taken.}
    \label{fig:threshold}
\end{figure}

We find that the threshold of this decoder varies between $p_c = 0.09$ and $p_c=0.14$, depending on the value of the bias, $\zeta_p = p_Z/p_Y$. The threshold is defined as the critical value $p_c = p_Y+p_Z$ such that below $p_c$, the chance of logical failure vanishes as the system size increases towards the thermodynamic limit. A plot of the threshold versus bias is shown in Fig.~\ref{fig:threshold}. There is a small improvement of about 5\% as bias increases, approximately linear in $1/\zeta_p$.

It is interesting that this improvement occurs in a decoder that does not use information about the bias at all. Rather, the improvement is coming from the different structure of the error syndrome at higher bias. This may be due to the fact that string-like errors become less likely as bias increases, leading to more syndromes for errors of the same weight.

\subsection{Results}
\begin{figure}[t!]
    \centering
    \includegraphics[width=\columnwidth]{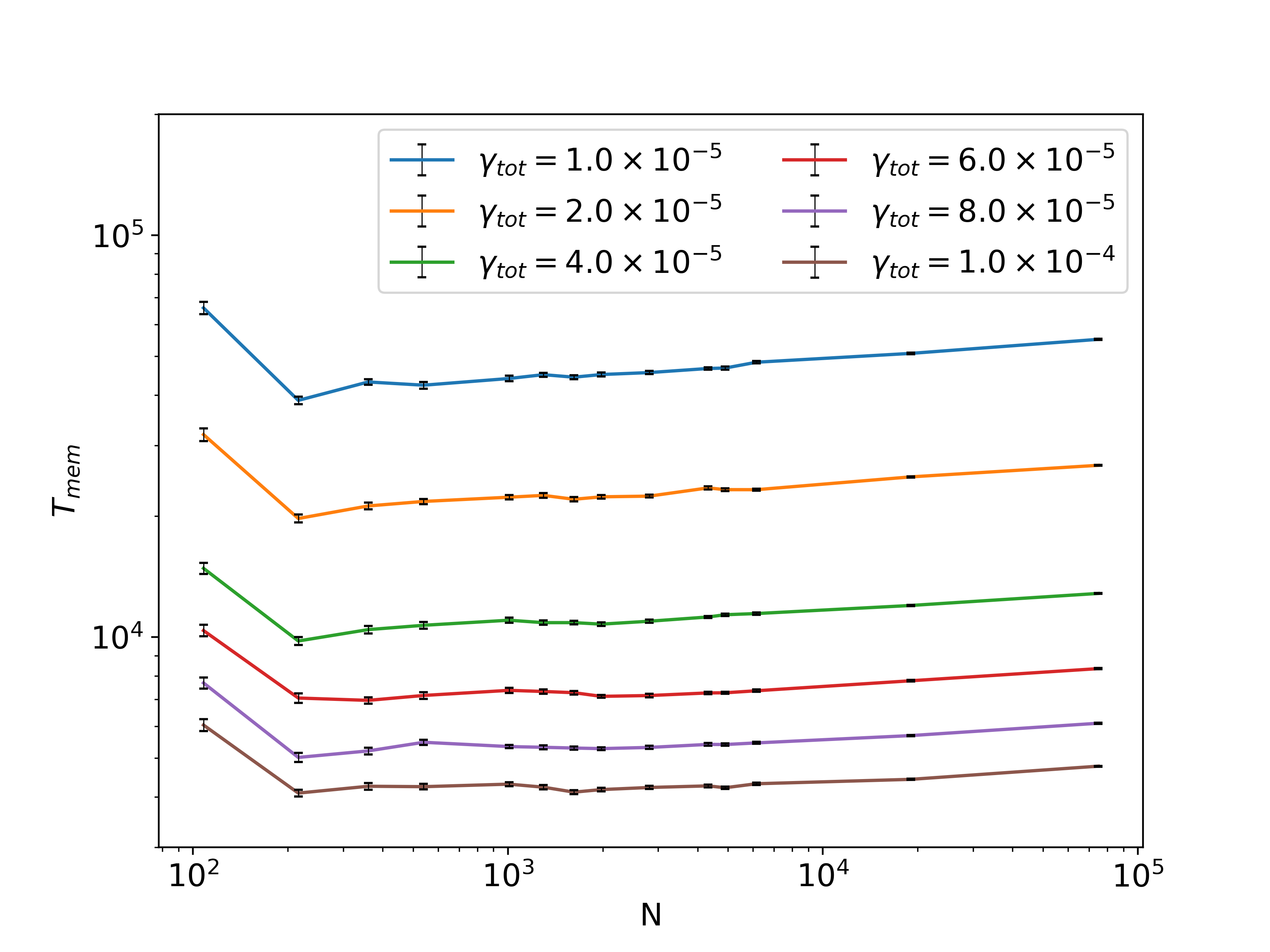}
    \caption{Memory time vs system size at bias $\zeta = 100$.}
    \label{fig:memvssizebias}
\end{figure}

The lack of partial self correction with finite bias is immediately apparent in Fig.~\ref{fig:memvssizebias}. We repeat the same measurements as in Fig.~\ref{fig:memvssystemsize}, but with a bias of $\zeta = 100$. Instead of finding partial self-correction, we find that memory time changes very little with system size. While there is a small decrease at the start, followed by an even smaller increase, these effects are many orders of magnitude smaller than the self-correcting behavior seen at infinite bias.

The steep initial decline of the curves in Fig.~\ref{fig:memvssizebias} is evidence that a very small number of $Y$ errors causes a significant decrease in memory time. To see this, we estimate $T_Y$  (Eq.~\ref{eq:TY}), the median time before the first $Y$ error. At the smallest system size of 6$\times$9, we have $T_Y \approx 0.64/\gamma_{tot}$. This value ends up being slightly larger than the measured memory time, which means that some simulations must occur without $Y$ errors, and thus match the infinite bias results. For any larger system size, however, $T_Y$ is much smaller than the measured memory time, so that $Y$ errors occur with high probability in every simulation run. 

The fact that 6$\times$9 is special is the result of our choice of $\zeta=100$. For any system size, we can always choose $\zeta$ such that $T_Y$ is high enough that it is highly improbable that any $Y$ errors occur before memory failure. However, as mentioned earlier, $\zeta$ must grow larger and larger with system size to ensure this.

\begin{figure}[t]
    \centering
    \includegraphics[width=\columnwidth]{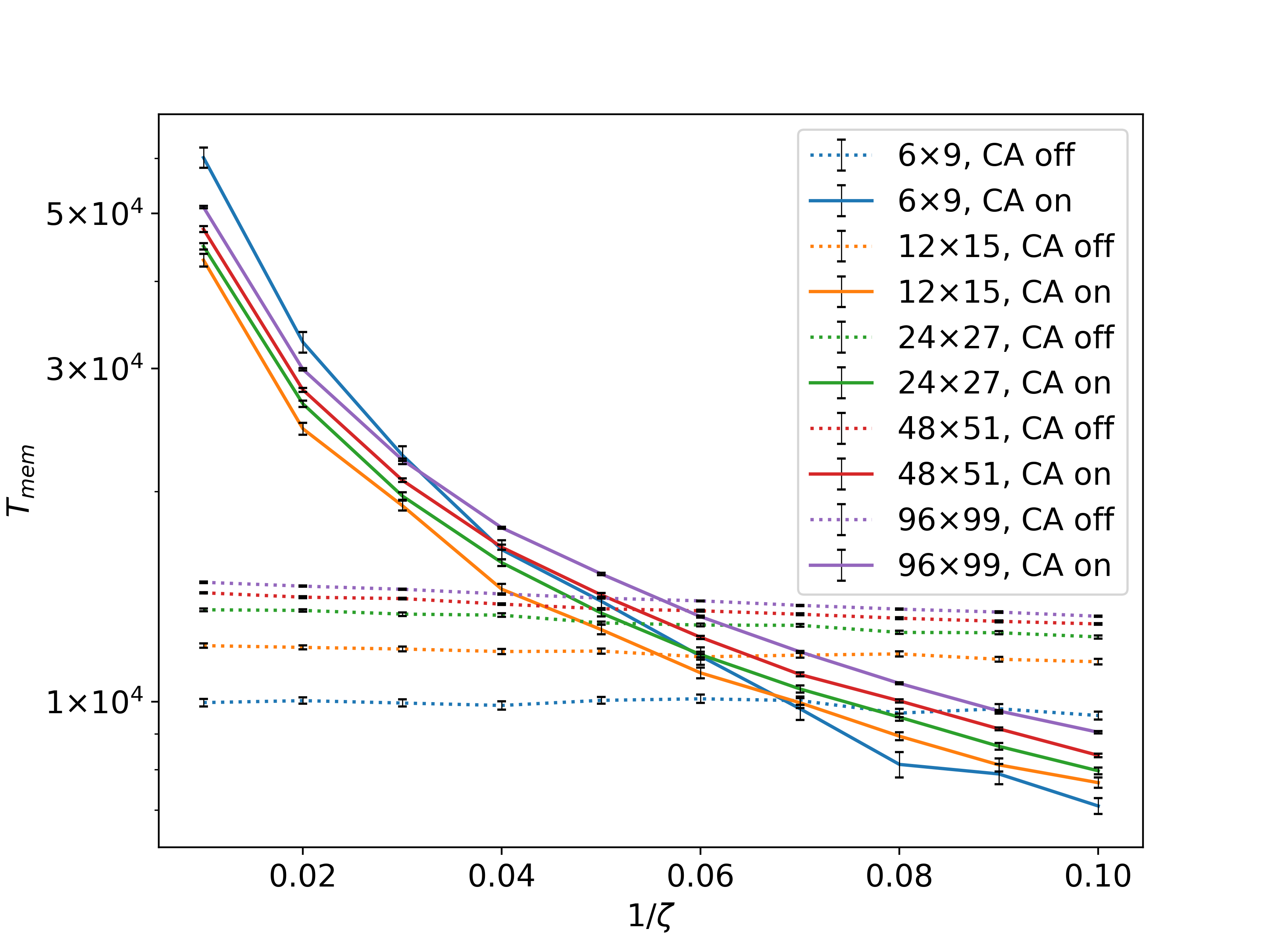}
    \caption{Comparison of memory times at values of the bias ranging from $\zeta = 10$ to $\zeta=100$, for total error rate of $\gamma_{tot}=10^{-5}$. Memory times without the cellular automaton rule are indicated with dotted lines.}
    \label{fig:biascompare}
\end{figure}

This property is illustrated more clearly in Fig.~\ref{fig:biascompare}. Here, we compare memory times at different values of the bias, holding total error rate fixed at $\gamma_{tot}=10^{-5}$. At the leftmost point in the figure, where $\zeta=100$, the system size with the highest memory time is 6$\times$9. However, as $\zeta$ falls to 10, the memory time of 6$\times$9 becomes worse than all of the other sizes, as $T_Y$ falls by about a factor of 10.

In Fig.~\ref{fig:biascompare}, we also plot the memory times of the XYZ color code where no cellular decoder is used, and compare them with the cases where we do use the cellular automaton decoder. Without a local decoder, errors simply accumulate until a logical error occurs. For large enough bias, it is clear that the cellular automaton still increases the memory time. For $\zeta > 25$, using the cellular automaton decoder outperforms the case where we use no local decoding for all system sizes shown. While this increase in memory time does not vary appreciably with system size, it does so substantially with bias. For architectures with a large but finite bias, it may then be efficient to use the local cellular automaton decoder to reduce the frequency of global decoding steps.

To summarize, we find that finite bias has a major destructive effect on memory time, removing most of the improvement with system size that we saw in Sec.~\ref{sec:infiniteresults}. The probability of memory failure increases steeply after just one $Y$ error occurs. However, even with these strong effects, the cellular automaton decoder still has improved memory time over the case without a local decoder for a reasonably wide range of biases and system sizes.

The extent to which these results depend on the decoder used, rather than the code itself, is still somewhat inconclusive. From Fig.~\ref{fig:threshold}, it is clear that the renormalization-group decoder has a far worse threshold than our exact decoder, even at very high bias. The renormalization-group decoder, however, is a general-purpose decoder with a threshold that is often exceeded by decoders tailor-made to specific codes. Furthermore, it has been previously shown in other codes that decoders exist with thresholds asymptotically approaching 50\% as bias approaches infinity\cite{bonilla2021}. It is unclear whether such an efficient decoder exists for the XYZ color code, and whether it would result in significant improvement of memory times.

\section{Conclusion}
\label{sec:conclusion}

To summarize, we have investigated error correction with the XYZ color code, a variant of the CSS color code whose physical qubits are locally rotated, undergoing errors from a biased noise model using a cellular automaton decoder. Under an infinitely biased noise model, logical errors have a fractal support, and as such, excitations have constrained dynamics akin to those of a type-II fracton model. Our simulations show that, at infinite bias, our local decoder reproduces the signatures of a partially self-correcting memory, namely, a memory time that diverges with system size, up to a critical system size. Therefore, in the limit of very high bias, with low error rates and modest system sizes, we expect that the code requires less intervention from a global decoding system as we scale the code distance. This may be valuable in the situation where the bandwidth between quantum hardware and classical control systems are limited. Although we were not able to reproduce the same partially self-correcting phenomena at finite biases using our local decoder, we have observed modest improvements in memory time for realistic biases we might expect to find with real hardware, of the order of $ \zeta \sim 100 $. Again then, it may be valuable to use a cellular automaton decoder in this regime to reduce the communication demands required by a global decoder.

The discovery of a self-correcting quantum memory will furnish us with a stabilizer code that can be decoded locally, thereby minimizing the communication demands for decoding. However, the dimensionality of known self-correcting memories means they are impractical for realization. As an interim, we can find partially self-correcting codes in lower-dimensional systems. Here we have witnessed partial self correction in a two-dimensional system in the limit of very high noise bias. 

An important question for future work is to understand the effects of imperfect measurements on the local decoder. In order to do this using our methodology, we must also realize the idealized continuous-time cellular automaton decoder as a discrete-time cellular automaton decoder, since a real-life circuit can only make measurements at a finite rate. In general, a continuous time cellular automaton is the limit of a discrete cellular automaton with a small time step, but it is possible that this limit is modified in the presence of measurement errors.

It will also be valuable to extend this partially self-correcting behavior to systems that experience finite bias. One approach to achieve this might be to investigate other local decoders. We might expect that we could discover better cellular automaton rules to this end that go beyond the intuition we have used to design our local decoder that is based on the physics of a finite-temperature environment. We might also consider using a cellular automaton decoder to decode three-dimensional partially self-correcting codes such as the cubic code. We may find that this code will serve as a practical alternative if classical communication is highly constrained, even where qubits experience a more generic noise model.

\vspace{0.2in}
\noindent{\it Acknowledgements} --- 
We are grateful for discussions with Stephen Bartlett, Arpit Dua, Markus Kesselring, Sam Smith, David Stephen and David Tuckett. JSM acknowledges support from the National Science Foundation Graduate Research Fellowship under Grant Number DGE-1656518. DJW and BJB are supported by the Australian Research Council via the Centre of Excellence in Engineered Quantum Systems (EQUS) project number CE170100009, and by the ARO under Grant Number: W911NF-21-1-0007. DJW also acknowledges support from the Simons foundation. 
Some of the computing for this project was performed on the Sherlock cluster. We would like to thank Stanford University and the Stanford Research Computing Center for providing computational resources and support that contributed to these research results. BJB changed affiliation to IBM Quantum during the preparation of this manuscript.

\bibliographystyle{plainnat}
\bibliography{bib}

\clearpage 

\onecolumngrid

\appendix

\section{Optimal System Sizes for Exact Decoding}
\label{app:optimalsize}
In this section, we prove that systems of size $3m$ by $3n$, where at least one of $m$ or $n$ is odd, always have two independent and inequivalent biased logical operators, which have extensive support. We then show that an infinite family of system sizes exist where these are the only biased logical operators.

\subsection{Existence of Logical Operators}
Recall that a biased logical operator is a logical operator made up of only one type of Pauli operator. Due to the decoupling of the stabilizer generators at infinite bias, a basis of biased logical operators exists such that each operator is supported on only one sublattice. We start by recalling that for logical operators in $\mathcal{P}_N^Z$ on the black sublattice, the following condition is required for each row:
\begin{align}\label{eq:approwrelation}
    R^{(j)}_i=R^{(j+1)}_i+R^{(j+1)}_{i+1},
\end{align}
where $R^{(j)}$ is a bitstring representing the support of the operator on row $j$. It can easily be checked that logical operators in $\mathcal{P}_N^X$ on the black sublattice satisfy the same relation. For logical operators on the white sublattice, in either $\mathcal{P}_N^Y$ or $\mathcal{P}_N^Z$, the relation is
\begin{align}\label{eq:rowrelationflipped}
    R^{(j)}_i=R^{(j-1)}_i+R^{(j-1)}_{i-1}.
\end{align}
This is simply Eq.~\ref{eq:approwrelation}, but flipped along both axes. Let us define the following notation to write down logical operators, using a matrix of bits $W$:
\begin{align}\label{eq:operatornotation}
    Z_a(W) = \prod_{i,j}Z_{i,j,a}^{W_{i,j}}.
\end{align}
For example,
\begin{align}
    Z_w\begin{pmatrix}
    1 0 \\
    0 1
    \end{pmatrix} = Z_{0,0,w}Z_{1,1,w}.
\end{align}
We use similar notation for $X$ operators, and operators on the black sublattice. An example of a matrix whose rows satisfy Eqs.~\ref{eq:approwrelation} and \ref{eq:rowrelationflipped} is:
\begin{align}\label{tileblock}
    \begin{pmatrix}
    101 \\
    011 \\
    110
    \end{pmatrix}.
\end{align}

\begin{figure}[h]
    \centering
    \includegraphics[width=\textwidth]{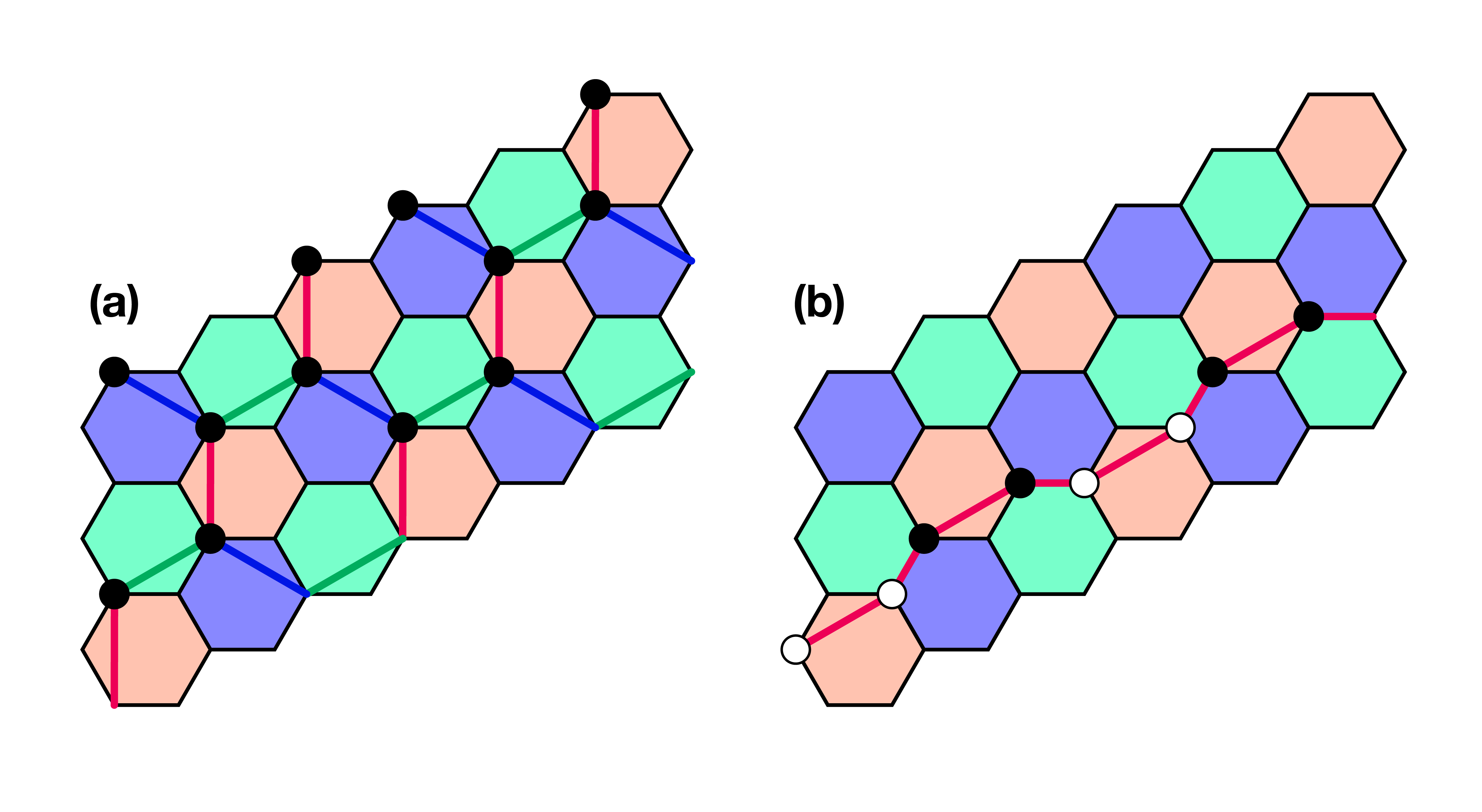}
    \caption{Logical operator of the XYZ color code on a $6\times 3$ lattice. White circles represent Pauli Y operators, and black circles are Pauli Z operators (also the white and black sublattices, respectively). Colored lines represent string nets in the CSS color code, see Ref.~\cite{bombin2006}. (a) The logical operator $M^{(Z)}_b$. (b) The string-like logical operator $\bar{Z}_{x,R}$.}
    \label{fig:logical}
\end{figure}

Since we require both dimensions of the system to be multiples of 3, we can tile this block to fill the entire system. We can also cycle the rows to get 3 different logical operators; however, one of them is the sum (mod 2) of the other two. Therefore, this makes two independent operators. One of these operators is shown on the hexagonal lattice in Fig.~\ref{fig:logical}.

\subsection{Commutation Relations}

We have so far shown that these operators commute with the stabilizers, but not that they are logical or independent. To do this, we find the commutation relations for a complete set of known logical operators for the XYZ color code. A simple basis of logical operators in the CSS color code is given by horizontal and vertical lines, of either $X$ or $Z$, on red or blue cells (the cells can always be colored, due to 3-coloring criterion of the lattice)\cite{bombin2006}. Let us denote these operators $\bar{Z}_{x,R},\bar{Z}_{x,B},\bar{Z}_{y,R},\bar{Z}_{y,B}, \bar{X}_{x,R},\bar{X}_{x,B},\bar{X}_{y,R},\bar{X}_{y,B}$. (Note that we use upper case $B$ to represent blue cells, and lower case $b$ to represent the black sublattice.) A pair of $\bar{Z}$ and $\bar{X}$ operators anticommute when they are of different colors and perpendicular to eachother.  Applying the local unitary to shift into the XYZ color code causes these to decompose into products of two string-like operators, each of a single Pauli type. For example:
\begin{align}\label{stringlogicals}
\bar{Z}_{x,R}=Z_b\begin{pmatrix}
000000 && \\ 
000000 && \hdots \\
011011 && \\
000000 && \\
\vdots && \ddots
\end{pmatrix}Y_w\begin{pmatrix}
000000 && \\ 
000000 && \hdots \\
101101 && \\
000000 && \\
\vdots && \ddots
\end{pmatrix},\quad \bar{Z}_{y,R}=Z_b\begin{pmatrix}
0000 &&\\
0100 &&\\
0100 && \hdots\\
0000 &&\\
0100 &&\\
0100 &&\\
\vdots && \ddots
\end{pmatrix}Y_w\begin{pmatrix}
1000 &&\\
0000 &&\\
1000 && \hdots\\
1000 &&\\
0000 &&\\
1000 &&\\
\vdots && \ddots
\end{pmatrix}.
\end{align}

The other logical operators can be derived from these two. We can cycle colors $R\rightarrow G \rightarrow B$ by shifting one column to the left. (This labeling of colors is not important, but the direction of the shifts must be consistent.) We can also take $\bar{Z}$ to $\bar{X}$ operators by taking $Z \rightarrow X$ and $Y \rightarrow Z$.

Next, let us denote our potential ``logical" operators as follows:
\begin{align}\label{potentiallogicals}
    L^{(X,Z)}_{(b,w)} = (X,Z)_{(b,w)}\begin{pmatrix}
    110 && \\
    101 && \hdots \\
    011 && \\
    \vdots && \ddots
    \end{pmatrix},\quad
    M^{(X,Z)}_{(b,w)} = (X,Z)_{(b,w)}\begin{pmatrix}
    101 && \\
    011 && \hdots \\
    110 && \\
    \vdots && \ddots
    \end{pmatrix}.
\end{align}
Here, the $\dots$ represents repeating the block. From Eq.~\ref{stringlogicals}, we see that every 3 columns of $\bar{Z}_{x,R}$ anticommute with one block each of $L^{(Z)}_{w}$, $M^{(Z)}_{w}$,  $L^{(X)}_w$, $M^{(X)}_b$, and $M^{(X)}_w$. The entire operators anticommute with each other if and only if the width is $3m$, where $m$ is odd. Likewise, $\bar{Z}_{y,R}$ anticommutes with $L^{(X)}_{b}$, $L^{(X)}_{w}$, and $L^{(Z)}_w$ if and only if the height is an odd multiple of 3.

Let us, for now, assume both dimensions are odd multiples of 3. We then calculate all of the commutation relations in a similar way to those for $\bar{Z}_{(x,y),R}$. Table \ref{tab:commutations} then gives the commutation relations for all $L,M$, and logical operators.
\begin{table}
    \centering
    \begin{tabular}{|c|c|c|c|c|c|c|c|c|}
        \hline
         & $\bar{Z}_{x,R}$ & $\bar{Z}_{y,R}$ & $\bar{Z}_{x,B}$ & $\bar{Z}_{y,B}$ & $\bar{X}_{x,R}$ & $\bar{X}_{y,R}$ & $\bar{X}_{x,B}$ & $\bar{X}_{y,B}$\\
        \hline
        $L^{(X)}_b$&+1&-1&-1&+1&+1&+1&+1&+1 \\
        \hline
        $L^{(X)}_w$&-1&-1&-1&+1&-1&-1&-1&+1 \\
        \hline
        $L^{(Z)}_b$&+1&+1&+1&+1&+1&-1&-1&+1 \\
        \hline
        $L^{(Z)}_w$&-1&-1&-1&+1&+1&+1&+1&+1 \\
        \hline
        $M^{(X)}_b$&-1&+1&-1&-1&+1&+1&+1&+1 \\
        \hline
        $M^{(X)}_w$&-1&+1&+1&-1&-1&+1&+1&-1 \\
        \hline
        $M^{(Z)}_b$&+1&+1&+1&+1&-1&+1&-1&-1 \\
        \hline
        $M^{(Z)}_w$&-1&+1&+1&-1&+1&+1&+1&+1 \\
        \hline
    \end{tabular}
    \caption{Commutation relations of logical operators.}
    \label{tab:commutations}
\end{table}
Based on these relations and the completeness of the basis of logical operators, we can determine, modulo products of stabilizers:

\begin{align}\label{bothoddlogicals}
\begin{split}
    L^{(X)}_b = \bar{X}_{x,B}\bar{X}_{y,R},\quad L^{(X)}_w = \bar{Z}_{y,R}\bar{Z}_{x,B}\bar{Z}_{y,B}\bar{X}_{y,R}\bar{X}_{x,B}\bar{X}_{y,B},\quad L^{(Z)}_b = \bar{Z}_{x,B}\bar{Z}_{y,R},\quad L^{(Z)}_w = \bar{X}_{y,R}\bar{X}_{x,B}\bar{X}_{y,B}, \\
    M^{(X)}_b = \bar{X}_{x,R}\bar{X}_{y,R}\bar{X}_{y,B}, \quad M^{(X)}_w = \bar{Z}_{x,R}\bar{Z}_{y,B}\bar{X}_{x,R}\bar{X}_{y,B},\quad M^{(Z)}_b = \bar{Z}_{y,R}\bar{Z}_{x,R}\bar{Z}_{y,B}\quad M^{(Z)}_w = \bar{X}_{x,R}\bar{X}_{y,B}.
\end{split}
\end{align}
As a check, we can see that all of the $(M,L)^{(X)}$ operators commute with eachother, as do the $(M,L)^{(Z)}$ operators. We can also check that each $(M,L)^{(X)}$ operator anticommutes with only one $(M,L)^{(Z)}$ operator, which is clear from Eq.~\ref{potentiallogicals} and the fact that both dimensions are odd.

In the case of both dimensions odd, therefore, the $L$ and $M$ operators generate the full Pauli group on 4 qubits. When restricted to biased noise (either $X$ or $Z$), these operators generate a maximal commuting subgroup.

What about when one or both dimensions are even? First, in the case of both dimensions, the $M$ and $L$ operators commute with every stringlike operator, and are therefore trivial. However, one can check that there is a further set of logicals generated by $6\times 6$ blocks, which can tile the system in this case. We will not discuss the even case further here.

The other case is when one dimension is even and the other is odd. As an example, say that the width is even, but the height is odd. In this case, all of the commutation relations involving $\bar{Z}_{x,R},\bar{Z}_{x,B},\bar{X}_{x,R},$ and $\bar{X}_{x,B}$ become +1. Then, Eq.~\ref{bothoddlogicals} becomes

\begin{align}\label{evenoddlogicals}
\begin{split}
    L^{(X)}_b = \bar{X}_{x,B},\quad L^{(X)}_w = \bar{Z}_{x,B}\bar{X}_{x,B},\quad L^{(Z)}_b = \bar{Z}_{x,B},\quad L^{(Z)}_w = \bar{X}_{x,B}, \\
    M^{(X)}_b = \bar{X}_{x,R}, \quad M^{(X)}_w = \bar{Z}_{x,R}\bar{X}_{x,R},\quad M^{(Z)}_b = \bar{Z}_{x,R},\quad M^{(Z)}_w = \bar{X}_{x,R}.
\end{split}
\end{align}

Again, a maximal commuting sector of the logical Pauli group is generated. However, in this case, 100$\%$ biased noise of both $X$ or $Z$ can only act on the same logical sector. Therefore, the $M$ and $L$ operators are not complete when one side has even length. This latter fact can be easily checked with Eq.~\ref{potentiallogicals}, and the fact that there are an even number of blocks in the system. The same holds when the width is odd, and the height is even.

\subsection{Other Logical Operators}

We now ask whether there are further logical operators besides those given by Eq.~\ref{tileblock}, when restricted to biased noise. We will look for independent combinations of all $X$ or all $Z$'s not given by blocks of Eq.~\ref{tileblock}. Note that these may not be independent when we quotient by products of stabilizers. However, we are still interested in finding as few of these logical operators as possible to simplify our calculations.

We can answer this question with results from Ref.~\cite{martin1984}. Before we begin, we note that most of the results in that reference are for rule 90. However, rule 90 on a string of length $2n$ is equivalent to two copies of rule 102 on a string of length $n$.

The important result we need is given by Lemma 3.4 of Ref.~\cite{martin1984}, which states that the lengths of all cycles divide the length of a cycle reached with an initial configuration that has a single 1 (for example, 100000...). Note that this initial configuration is not part of a cycle, but it must eventually reach a cycle due to the finiteness of the system.

This gives a simple construction for system sizes that only allow the operators with cycle length 3. First, choose a width $L=3n$. Next, find the length $\Pi_L$ of the cycle reached from the bitstring with a single 1. Finally, let the system height be $H=3p$, where $p$ is a prime that does not divide $\Pi_L$. This guarantees $GCD(H,\Pi_L)=3$, so that the only cycles are of length 3. Another construction is to take $L=3(2^n)$ and $H=3(2^n+1)$. In this case $\Pi_L$ can be shown to be $3(2^n)$ (see Fig. \ref{fig:CAevolutions} (b) for an example), so only length 3 cycles are allowed here as well.

In fact, constraining the logical operators to have periodicity 3 in the $y$-direction also forces them to have periodicity 3 in the $x$-direction, which forces them to be of the form of Eq.~\ref{tileblock}. To see this, we again denote the bit at position $i,j$ of a logical operator as $R^{(j)}_{i}$. Then, $GCD(h,l)=3$ implies that $R^{(j)}_{i}=R^{(j-3)}_{i}$. But we can also use Eq.~\ref{eq:rowrelationflipped} to show:

\begin{align}
R^{(j)}_{i}=R^{(j-3)}_{i}+R^{(j-3)}_{i-1}+R^{(j-3)}_{i-2}+R^{(j-3)}_{i-3}.
\end{align}

Similarly, we have

\begin{align}
R^{(j)}_{i-1}=R^{(j-3)}_{i-1}+R^{(j-3)}_{i-2}+R^{(j-3)}_{i-3}+R^{(j-3)}_{i-4}.
\end{align}

Using the fact that $R^{(j)}_{i}=R^{(j-3)}_{i}$, and adding the equations together, we find

\begin{align}
R^{(j)}_{i-1} = R^{(j)}_{i-4}.
\end{align}

Therefore, the logical operator must have periodicity 3 in the $x$ direction. By checking all 8 cases, we can see that the only nontrivial possibilities are of the form in Eq.~\ref{tileblock}.

\section{Optimal Decoder Threshold}
\label{app:proofthreshold}

In this section, we prove the following lemma, from which the 50\% threshold of the XYZ color code at infinite bias follows. We restate the lemma here.

\textbf{Lemma 1}: consider a family of error correction codes $\mathcal{C}$, parameterized by system size $N$. Let the error channel be i.i.d. and infinitely biased. Define $\mathcal{L}(N)$ to be the set of biased logical operators. If $|\mathcal{L}(N)|$ is constant with $N$, and if every error in $\mathcal{L}(N)$ has support polynomial in $N$, then a maximum likelihood decoder for $\mathcal{C}$ has a 50\% threshold.

\textbf{Proof}: A maximum likelihood decoder fails for any error $E$ when there exists a logical error $L$ such that $LE$ is more likely than $E$.

\begin{align}\label{eq:originalpfail}
    P_{fail}(N)=\sum_{\substack{E\\P(LE) \geq P(E), L \in \mathcal{L}(N)}} P(E),
\end{align}

For lightness of notation, we keep the functional dependence on $N$ implicit for the remainder of the proof. This sum may be bounded from above:

\begin{align}
    P_{fail} \leq \sum_{L\in \mathcal{L}}\sum_{\substack{E\\P(LE)\geq P(E)}}P(E).
\end{align}

This overcounts Eq.~\ref{eq:originalpfail}, because there may be some errors for which there is more than one logical error $L$ which satisfies $P(LE) \geq P(E)$.

We now define the error $E$ as the product $E=E_L E_{\bar{L}}$, where $E_L$ is the error restricted to the support of $L$, and $E_{\bar{L}}$ is the error on the complement. Then

\begin{align}
    P(E) = P(E_L)P(E_{\bar{L}}).
\end{align}

Let us suppose $p_Z < \frac{1}{2}$. Therefore, if $P(LE) \geq P(E)$, 

\begin{align}
    |E_L| \geq |L| / 2,
\end{align}

Since at least half of the errors in $L$ must cancel errors in $E$ to reduce the total support. Now, since every $L$ in $\mathcal{L}$ has support growing polynomially in system size $N$, there exist positive constants $K_L$, $\alpha_L$ such that if $P(LE) \geq P(E)$,

\begin{align}
    |E_L| \geq K_L N^{\alpha_L} / 2.
\end{align}

Therefore, 

\begin{align}
    P_{fail} \leq \sum_{L\in\mathcal{L}} \sum_{E_{\bar{L}}} P(E_{\bar{L}}) \sum_{E_L, |E_L|>K_L N^{\alpha_L} / 2} P(E_L) \leq \sum_{L\in\mathcal{L}} \sum_{E_L, |E_L|>K_L N^{\alpha_L} / 2} P(E_L).
\end{align}

This can be expressed using the cumulative distribution function of the binomial distribution:

\begin{align}
    P_{fail} \leq \sum_{L\in\mathcal{L}} \sum_{i=K_L N^{\alpha_L} / 2}^{K_L N^{\alpha_L}}p_Z^{i}(1-p_Z)^{N-i} = \sum_{L\in\mathcal{L}}F(K_L N^{\alpha_L}/2 ; K_L N^{\alpha_L}, 1-p_Z),
\end{align}

Where $F(k;n,p)$ is the cumulative distribution function of the binomial distribution with $n$ trials and $k$ or fewer successes. Due to Hoeffding's inequality\cite{hoeffding1963}, this can be bounded by

\begin{align}
    F(k;n,p)\leq \exp \left(-2n\left(p-\frac{k}{n}\right)^2\right).
\end{align}

We then have

\begin{align}
    P_{fail} \leq \sum_{L\in\mathcal{L}} \exp\left(-2K_LN_L^{\alpha_L}\left(\frac{1}{2}-p_Z\right)^2\right).
\end{align}

Since there are a constant number of $L$ in $\mathcal{L}$, this sum vanishes as $N \rightarrow \infty$, when $p_Z < \frac{1}{2}$. This proves the lemma.

\end{document}